\documentstyle[aps,prc,epsf,preprint,tighten]{revtex}

\def\beq{\begin{equation}}
\def\eeq{\end{equation}}
\def\bea{\begin{eqnarray}}
\def\eea{\end{eqnarray}} 
\def\beqa{\begin{equation}\begin{array}{l}}
\def\eeqa{\end{array}\end{equation}}
\def\eqlab#1{\label{eq:#1}}
\def\figlab#1{\label{fig:#1}}
\def\tablab#1{\label{tab:#1}}
\def\seclab#1{\label{sec:#1}}
\def\eref#1{(\ref{eq:#1})}
\def\eqref#1{eq.~(\ref{eq:#1})}
\def\Eqref#1{Eq.~(\ref{eq:#1})}

\def\Figref#1{Fig.~\ref{fig:#1}}
\def\tabref#1{\ref{tab:#1}}

\def\secref#1{Sec.~\ref{sec:#1}}
\def\sla#1{#1 \!\!\!\! \slash}

\def\slap{p \!\!\! \slash}

\def\half{\mbox{\small{$\frac{1}{2}$}}}
\def\thalf{\mbox{\small{$\frac{3}{2}$}}}
\def\quarter{\mbox{\small{$\frac{1}{4}$}}}
\def\third{\mbox{\small{$\frac{1}{3}$}}}

\def\barr{\left(\begin{array}{c}}
\def\earr{\end{array}\right)}
\def\bmat{\left(\begin{array}{cc}}
\def\emat{\end{array}\right)}
\def\al{\alpha}
\def\be{\beta}
\def\ga{\gamma} \def\Ga{{\it\Gamma}}
\def\de{\delta} \def\De{\Delta}
\def\veps{\varepsilon}  \def\eps{\epsilon}

\def\la{\lambda} \def\La{{\it\Lambda}}

\def\si{\sigma} \def\Si{{\it\Sigma}}
\def\th{\theta}  
\def\w{\omega}  
\def\vfi{\varphi}

\def\dd{{\rm d}}
\def\pa{\partial}

\def\BK#1#2{{\it #1}, #2}         
\def\CF#1#2#3#4{#1 {\bf #2}, #4 (#3)}  
\def\ibid {{\it ibid.}}

\def\ann {Ann.~Phys.~(NY)}
\def\fbs {Few-Body~Systems}
\def\ijmp {Int.~J. Mod. Phys.}

\def\jpg {J.~Phys.~G}

\def\ncim {Nuovo Cimento}
\def\np {Nucl.~Phys.}
\def\prev {Phys. Rev.}
\def\prc {Phys. Rev.~C}
\def\prd {Phys. Rev.~D}
\def\plett {Phys.~Lett.}
\def\plb {Phys.~Lett.}
\def\prl {Phys. Rev.~Lett.}

\def\ie{{i.e., }}

\def\ni{\noindent}
\def\pa{\partial}

\def\no{\nonumber}

\def\rg{{\rm g}} \def\rp{{\rm p}}
 
\def\rq{q} \def\hq{\hat{q}}
\def\hk{\hat{k}}
\def\tk{\tilde{k}}
\def\rN{{\rm N}} 

\def\tq{\tilde{q}} \def\tl{\tilde{l}}
 
\def\cO{\mathcal{O}}

\def\lag{{\mathcal L}}

\def\MM{{\mathcal M}}

\def\mathscr{\mathcal}
\def\N{N}
\def\bGa{{\bf\Gamma}}
\def\bS{{\bf S}}

\def\vp{\vec{p}}

\def\vq{\vec{q}}
\def\pn{$\pi N$ }
\def\arctg{{\rm arctg}}
\def\re{\mbox{Re}}

\def\3d{3D}
\def\CMS{CMS}
\def\becol{}
\def\ecol{}

\begin{document}

\title{
Pion-nucleon interaction in a covariant
hadron-exchange model}
\author{V.~Pascalutsa$^{1,2}$ and J.~A.~Tjon$^{1,3}$}
\address{$^{1}$ Institute for Theoretical Physics, 
University of Utrecht,
Princetonplein 5, NL-3584 CC Utrecht, The Netherlands}
\address{$^{2}$ National Institute for Nuclear and High
Energy Physics (NIKHEF),
 P.O.\ Box 41882, NL-1009 DB Amsterdam, 
The Netherlands}
\address{$^{3}$KVI, University of Groningen, 9747 AA Groningen, 
The Netherlands}
\date{\today}
\preprint{\framebox{NIKHEF 99-021}}

\maketitle
\thispagestyle{empty}

\begin{abstract}
We develop a relativistic covariant and unitary description of
the pion-nucleon interaction in a hadron-exchange model. 
The model is based on the solution of a dimensionally reduced
(quasipotential) Bethe-Salpeter  equation for the partial-wave
off-shell \pn scattering amplitudes with the potential
consisting of the field-theoretical $s$- and $u$-channel 
nucleon, Roper, Delta, $D_{13}$, $S_{11}$ exchanges, 
and the $t$-channel $\rho$ and $\sigma$ meson exchanges. 
The contributions of the spin-3/2 Delta and $D_{13}$ resonances  
are treated within the Rarita-Schwinger formalism and
different forms of the $\pi N\De$ vertex are investigated.
The free parameters of the model are fitted to the $\pi N$ 
phase-shift data of the KH80 and SM95 partial-wave analyses 
in the region up to 600 MeV pion kinetic energy.  The resulting
on-shell solution provides a good description of the \pn 
scattering lengths, as well as the energy behavior of the S, P,
and D partial waves. The sensitivity of the phase shifts on 
various model-dependent effects is examined.
\end{abstract}

\pacs{13.75.Gx, 21.45.+v, 12.39.Pn, 11.10.St}

\becol
\section{Introduction}

The \pn interaction received much attention in the past both 
theoretically and experimentally in view of its fundamental nature 
(early literature can be found in Refs.~\cite{early,Hoh83}). 
Current theoretical interest is triggered by the
experimental programs being carried out at NIKHEF, MAMI, 
TJNAF and other intermediate-energy facilities
with the purpose to understand the structure of hadrons
and their interaction in the confinement region of QCD. 
To extract most physics from the new high-precision
measurements a reliable and accurate knowledge of \pn and
$NN$ interaction is required.
Highly successful attempts have been made in describing these 
interactions  in terms of {\it hadronic degrees of freedom} over a 
wide energy region. In particular, the {\it relativistic 
one-boson-exchange models} were successfully applied
to the description of $NN$ interaction, and especially
during the past decade 
this theoretical framework has been extended to the \pn system 
\cite{PeJ91,LYL91,GrS93,SDH94,HYL94,SaL96,LaA98,PaT98}. 
Such an extension gives one a capability to study a broad class
of reactions, including
pion scattering and production on light nuclei in a self-consistent
framework. In this paper we report on 
a relativistic covariant $\pi N$ model using the quasipotential
approach and based on an effective interaction characterized 
by a hadron-exchange potential (some of our results have already 
been briefly reported \cite{PaT98,PaT98a}).

Although the underlying dynamics of the $\pi N$ interaction
is nowadays believed to be governed by QCD, it is practically
impossible to resolve it fully in terms of quarks and gluons 
because of the confinement problem. Much of the present
understanding of the \pn physics at low and intermediate energies
remains to be based on dispersion relations \cite{Hoh83} and
effective chiral Lagrangians \cite{GeL60,Wei67} in term of the hadron
degrees of freedom.

The chiral pion-nucleon Lagrangians are usually extended in two
ways: first, by including higher-mass states, such as $\rho$-meson,
$\De$-isobar, etc.; secondly, by including the higher-derivative
terms. Both ways are necessary to extend such
a phenomenological description to higher energies: The contributions 
due to higher-mass states have a clear physical significance, while
the higher-derivative terms are needed  to examine
the effect of unknown short-range physics. 
The higher-derivative terms play, for instance, a crucial role
in the renormalization program of chiral perturbation
theory (ChPT). In the hadron-exchange models, 
a similar role is played
by the ``strong form factors'' which are included 
in the effective Lagrangian to model the short distance behavior
of the potential. Both ChPT and hadron-exchange models thus 
begin from a similar ``extended'' chiral Lagrangian but the
approach to calculating the $\pi N$ scattering amplitude
is somewhat different. In ChPT one usually performs a perturbative
field-theoretic calculation maintaining crossing-symmetry
and exact agreement with the soft-pion theorems.
(For the development of ChPT in application to 
the \pn scattering see Ref.~\cite{eft}.)
In the hadron-exchange approach one uses the effective Lagrangian
to construct the potential which then is resumed via
a scattering equation. In this way crossing symmetry is given
up in favor of exact unitarity in a given channel space, and 
possibility of studying nonperturbative phenomena such as 
dynamical resonances. 

In defining a hadron-exchange model one usually specifies
three ingredients: effective Lagrangian, potential, and
the scattering equation. These ingredients are interrelated 
in quantum field theory, where one must solve the 
Bethe-Salpeter (BS) equation and has a well-determined procedure
for computing its kernel from a given Lagrangian.
The BS kernel consists of all the
irreducible graphs and hence can be computed only
perturbatively. In this work we shall take the potential to be given
by the tree-level graphs, although we do not claim that
the used Lagrangian justify any perturbative expansion.
On the other hand, the resulting approximation transparently
relates to the usual quantum-mechanical picture where
the scattering problem is given by a Lippmann-Schwinger
type of equation for one-particle exchange potentials.
Therefore, one might prefer to view this approach
as relativistic quantum-mechanical one rather than
some ``nonsystematic'' truncation of QFT. 

The four-dimensional BS equation for the
$\pi N$ system (with a one-particle-exchange potential)
has been solved by Nieland and Tjon \cite{NiT68}, and recently by
Lahiff and Afnan \cite{LaA98} in a more realistic setup. 
Models \cite{PeJ91,GrS93,HYL94,PaT98} exploit instead
various {\it quasipotential} (QP) equations, which can be obtained
by a three-dimensional reduction 
of the BS equation. The use of QP equations provides
a technical simplification of the problem, without destroying the
Lorentz invariance of the theory. It should be remarked, however,
that some of the QP equations can violate charge conjugation symmetry,
and because this symmetry is crucial for renormalizing
the positive- and negative-energy baryon poles in an equivalent way, 
the equations which violate it are less preferable. 
We will employ the
{\em equal-time} (ET) {\em equation} which preserves the 
full Lorentz covariance, including charge conjugation. This 
equation will be specified in the next section.

Apart from the technical simplifications, the QP approach 
can sometimes be motivated by physical arguments. For instance, 
while the four-dimensional BS equation for $t$-channel 
(meson-exchange) potential, \ie the {\it ladder} BS equation, has a  
wrong one-body limit, a number of QP equations with the proper limit
can be devised \cite{Grs69,Tod73,MaW87,Cj89}. In the $\pi N$ situation
the potential may, in addition to the $t$-channel meson exchanges, 
contain the $u$-channel baryon exchanges, which spoil
the standard one-body limit arguments \cite{PaT99a}.
The ordinary ET equation is shown to provide an optimal choice 
in the case when both $t$- and $u$-channel exchanges are present
in the force corresponding to the \pn situation\cite{PaT99a}. 

As for the effective Lagrangian, we use the
pseudovector $\pi NN$ coupling, and include $\si$, $\rho$
mesons, and all the relevant
(for the considered energy region) nucleon resonances
as explicit degrees of freedom. The precise form
of the Lagrangian and the potential is discussed in
Sec.~IV. An interesting aspect,
which comes in with the nonperturbative modelling, is that
for a sufficiently attractive potential the nucleon
resonances can be generated dynamically, as
quasi-bound states of the $\pi N$ system. In
models \cite{NiT68,ChL56} and quantum hadrodynamics \cite{SeW97}
the $\De$(1232) is described in this way.
Lahiff and Afnan \cite{LaA98} include the $\De$ explicitly,
but suggest that the Roper resonance can be of dynamical
origin. Gross and Surya \cite{GrS93} include the $\De$
and the Roper poles, but treat the $S_{11}$ resonance dynamically. 
In this paper we consider 
$P_{33}\,\De$(1232), $P_{11}\,N^\ast$(1450), 
$D_{13}\,N^\ast$(1520), and $S_{11}\,N^\ast$(1535) resonances. 
Within our model, these resonances are all of nondynamical origin,
i.e., are included explicitly via an effective Lagrangian description.
Of course, the dynamical effects will anyhow
contribute to the generation of the
resonances seen in the phase-shifts. Thus, an 
admixture of both ``elementary'' and 
``composite'' component constitutes the full result. 
Since the ``elementary''
fields corresponding to the resonances are included with real masses,
the dynamical contributions are fully responsible for generating
the width. Our model maintains the elastic $\pi N$ unitarity 
and therefore only the one-pion decay width of the resonances
is generated.  

As we have to deal with the spin-3/2 fields of resonances, such 
as that of the $\De$-isobar, we shall address here some of the 
problems of 
consistent formulation for relativistic higher-spin fields. Consistent
formulations for the {\it free} spin-3/2 field have of course been known
for a long time. The Rarita-Schwinger formalism \cite{RS41} 
based on the vector-spinor field representation became the most 
popular one.
The form of the free spin-3/2 action is uniquely (up to trivial
field redefinitions) constrained by requirements of Poincar\'e 
invariance and consistent degrees-of-freedom counting. The latter
requirement essentially means that the action must have
enough symmetries to kill off the unphysical lower-spin components,
and maintain only the physical $2s+1$ degrees-of-freedom of the theory. 
An arbitrarily constructed interaction may violate this consistency
and activate the unphysical degrees of freedom. This necessarily leads
to a number of pathologies, such as acausal 
propagations\cite{VZ69,Sin73}, 
inadmissible quantization\cite{JS61,Hag71}, violation of Poincar\'e 
invariance. 

The conventional $\pi N\De$ 
coupling\cite{Pec68,Nat71,Hoh83} is an example
of such pathological interactions. The contribution of
the unphysical spin-1/2 components appear in the scattering
amplitudes via the dependence on the so-called ``off-shell parameter''
and ``spin-1/2 backgrounds.'' On the other hand, a class
of consistent $\pi N\De$ and $\ga N \De$ couplings has recently 
been established \cite{Pas98,PT99}. Those are essentially 
all possible couplings that maintain the {\em gauge symmetry} of 
the free massless Rarita-Schwinger
action. In the present model we shall use the
leading ``gauge-invariant'' $\pi N\De$ coupling, which has the 
same non-relativistic limit as the conventional one. 
In \secref{solution} we study the  differences between the conventional
and the gauge-invariant $\pi N\De$ coupling at the tree-level
$\De$-exchange contributions. The largest differences are 
seen first of all in the spin-1/2 partial waves, where
the conventional coupling gives the background contributions
verse no contribution from the consistent couplings.

The parameters of the effective Lagrangian, 
including the form factor masses, form the set of model parameters. 
Some of them, such as the $\pi NN$ coupling constant, the nucleon,
and the meson masses, are very well-determined elsewhere and therefore
are kept fixed during the fits.
The others are fitted to give the best 
agreement with the \pn partial-wave
analyses \cite{ASW95,KH80,KA86}. 
The complete model provides an accurate description of the
$S$- and $P$-wave scattering lengths 
($\chi^2/{\rm data point} \simeq 1.4$), 
as well as the energy behavior of the $S$-, $P$-, and 
some of the $D$-wave phase-shifts up to 600 MeV pion
lab kinetic energy.

In the next section we describe the covariant quasipotential 
equation for the off-shell \pn amplitudes, and its 
partial-wave decomposition. 
This will specify the equation solved in the model.
In \secref{pot} we briefly discuss the effective Lagrangian
used to read off the tree-level potential---the driving force of
the equation. 
The renormalization procedure to treat the s-channel poles 
is described in \secref{renorm}. 
In \secref{solution} we analyze the effects of the
different exchange contributions in the low-energy \pn data,
and then present the results of the complete model 
emphasizing the effect of the rescattering contributions.
Some discussion and concluding remarks are given in \secref{concl}. 
Finally, various appendices contain some technical details of
the analysis.
In Appendix A we summarize the conventions. Appendix B shows
some details of the partial-wave and isospin decomposition of the 
off-shell \pn
amplitudes. Appendix C provides explicit
expressions for various hadron-exchange contributions to 
the off-shell \pn potential. Amplitudes for higher-spin baryon
exchange are discussed in Appendix D.

\section{Quasipotential approach}
\seclab{qp}

The fully off-shell relativistic $\pi N$ scattering amplitude 
in the space of the nucleon helicity spinors is described by
sixteen scalar amplitudes: one for each combination of the
helicity and $\rho$-spin of the initial and final nucleon.
Parity conservation reduces the number of independent scalar 
amplitudes to eight. As a suitable covariant representation which
expresses the off-shell amplitude in terms of eight invariants
we choose the following:
\beq
\eqlab{offt}
T_{\la'\la}^{\rho'\rho}(p',k';p,k) = \bar{u}_{\la'}^{\rho'} (\vp')\,
 (1, \slap') \left[ \bmat A_{11} &  A_{12}\\ 
  A_{21}  & A_{22}  \emat + \sla{P} \bmat B_{11} &  B_{12}\\ 
  B_{21}  & B_{22}   \emat
\right] \barr 1\\ 
  \slap \earr \,u_{\la}^{\rho}(\vp), 
\eeq
where $A_{ij}$ and $B_{ij}$ are the eight scalar functions of 
invariants formed by the momenta, \ie $p^2$, $p\cdot p'$, etc.
The four-momenta of the initial (final)
nucleon and pion are given by $p$ and $k$ ($p'$ and $k'$), respectively, 
while $P$ is the conserved total
four-momentum of the system:
$$
P=p+k=p'+k'\, .
$$
Due to the momentum conservation only three of the
external momenta are independent, below we usually work with $p$,
$p'$ and $P$.
Furthermore, $u_\la^\rho$ are the 
nucleon helicity spinors, where
$\la=\pm\half$ and $\rho=\pm 1$ 
($\la'$ and $\rho'$) are the initial (final) 
helicity and $\rho$-spin of the nucleon, respectively.

For the on-shell situation ($p^2={p'}^2=m_N^2$, $k^2={k'}^2=m_\pi^2$)
the amplitude reduces to the standard form \cite{CGLN}:
\beq
\eqlab{ont}
T_{\la'\la}(p',p;P) = \bar{u}_{\la'}^{(+)}(\vp')\,
\left[ \hat{A}(s,t) + \half\ga\cdot(k'+k)\,
\hat{B}(s,t)\right]\,u_{\la}^{(+)}(\vp), 
\eeq
[$s=P^2,\,t=(p-p')^2,\,u=(p-k')^2$ are the Mandelstam invariants].
We find from Eqs. \eref{offt} and \eref{ont}
\bea
\hat{B}&=& B_{11} + m_N(B_{12}+B_{21})+m_N^2 B_{22},\\
\hat{A}&=& m_N \hat{B} + A_{11} + m_N(A_{12}+A_{21})+m_N^2 A_{22}.
\eea

Our starting point for the $\pi N$ amplitude is the Bethe-Salpeter (BS) 
equation, schematically shown in \Figref{bsef},
\beq
\eqlab{bseq}
 T(p',p)=V(p',p)+ i\int\! \frac{\dd^4 q}{4\pi^3}\,V(p',q)\,G(q)
\,T(q,p),
\eeq
where $V$ is the potential, $G$ is the \pn propagator; 
the dependence on the total momentum $P$ is omitted.
If $q$ is the relative four-momentum of the intermediate $\pi N$ state, 
the $\pi N$ propagator takes the following form: 
\beq
G(q) = \frac{1}{(\be P-q)^2-m_\pi^2+i\veps}\,
\frac{(\al P + q)\cdot\ga + m_N}{(\al P + q)^2-m_N^2+i\veps},
\eeq
where 
\bea
&&\al\equiv\al(s)=p\cdot P/s=(s+m_N^2-m_\pi^2)/2s,\\
&&\be\equiv\be(s)=k\cdot P/s=(s-m_N^2+m_\pi^2)/2s.
\eea

In approximating the BS equation one
often simplifies the singularity structure of the kernel $VG$,
such that the temporal integration can easily be done.
This procedure is called {\it three-dimensional (3D) reduction}
while the resulting equation is a {\it quasipotential (QP) equation}.
For instance, in the reduction to the {\it spectator equation} 
\cite{Grs69,GrS93} all the poles of $V$
and the negative-energy pole of $G$ in the $q_0$ plane
are neglected.

As we have recently emphasized \cite{PaT98,PaT98a}, 
the danger of doing a 3D reduction via approximating the pole
structure is that the {\it charge conjugation symmetry}
can be destroyed. In particular, in our naive interpretation
of the spectator equation this symmetry is violated, essentially
because of an asymmetric treatment of the positive- and
negative-energy states. Gross has recently presented
an interpretation of the spectator prescription
which is consistent with the 
charge conjugation symmetry \cite{Grs99}.

The {\it equal-time} (ET) 
reductions (see, e.g. \cite{tlh}), such as
Salpeter's {\it instantaneous approximation} \cite{Sal52},
 preserve charge conjugation symmetry.
In these reductions one effectively removes
the $q_0$ poles from the potential while  
treating  exactly the poles of the two-particle propagator $G$.
To remove the potential poles one fixes the relative-energy 
variable $q_0$ in some way. 
Most frequently the constraint $q_0=0$, or its 
covariantized form: $P\cdot q = 0$, is used. 

We will be using the ET type of approach.
To implement the
constraint $P\cdot q=0$, we may impose the condition
that the interaction is insensitive to the off-shellness along
the direction defined by an unit four-vector $n_\mu$. 
For the two-body
case this means that $V$ and $T$ entering the scattering equation
depend on the
projections of the relative four-vectors onto a 3D hyperplane
orthogonal to $n_\mu$. Defining the projection operator,
\beq
\eqlab{etproj}
O_{\mu\nu}=\rg_{\mu\nu}-n_\mu n_\nu, 
\eeq
we write the corresponding equation as follows:
\beq
\eqlab{compact}
T(\tl',\tl)=V(\tl',\tl)+ i\int\,\frac{\dd^4 q}{4\pi^3} \,
 V(\tl',\tq)\, G(q)\, T(\tq,\tl),
\eeq
where $l,\,l',\,q$ are the {\it relative} momenta of the initial,
final, and intermediate
\pn state, respectively; $\tl_{\mu} = O_{\mu\nu} l^{\nu}$,   
and similarly for $\tl',\,\tq$.

Equation \eref{compact} is manifestly covariant.
On the other hand, it can easily be reduced to 
the 3D form. 
For instance, we can choose the frame where 
$n=(1,0,0,0)$, and therefore $V$ and $T$ are independent
of the 0-th component of relative momenta (since any scalar product
will depend only on the spatial components, e.g., $\tq\cdot\ga=-
\vec{q}\cdot\vec{\ga}$). The integration over $q_0$ in \Eqref{compact}
can now be readily done leading to the 3D equation. 
To prevent the dependence of the $S$-matrix on $n$, one may choose 
$n$ along some physical four-momentum, for instance along the
total momentum: $n_\mu=P_\mu/\sqrt{P^2}.$ Then the
reduction is possible in the center-of-mass
system (CMS) where $P=(P_0,0,0,0)$.  

In the ET approach, the two-particle propagator is sometimes modified
to include approximately the crossed graphs \cite{MaW87,TiT94},
thus providing the correct one-body limit of the equation
in the case of $t$-channel type of potential. We however
do not apply such modifications here, because
they actually worsen the predictions for the \pn case
where the $u$-channel exchanges are present \cite{PaT99a}.

Because of rotational invariance and parity conservation
it is convenient to partial-wave decompose
Eqs. \eref{bseq} and \eref{compact} (see Appendix B).
Let
\beq
T_{\la' \la}^{\rho' \rho}(p',p)=\sum\nolimits_J (J+\half) D^J_{\la'
\la}(\Omega_{p'p})\, T_{\la' \la}^{J\rho' \rho}(p_0',\rp',p_0,\rp; P_0), 
\eeq
where
$\Omega_{p'p}$ is the solid angle between $\vec{p}'$ and $\vec{p}$.
Furthemore, in the CMS, 
using \Eqref{offt} and the Dirac equation
\beq
(\slap -\rho m_N)\, u_{\la}^\rho (\vp)=(p_0 -\rho E_p)\ga_0\,
u_{\la}^\rho (\vp),
\eeq
the off-shell amplitudes can be written as
\beq
\eqlab{defparity}
T_{\la'\la}^{\rho'\rho}(p',p;P_0) = \bar{u}_{\la'}^{\rho'} (\vp')\,
 \left[ \ga_+\, T_+^{\rho'\rho}(p',p;P) + \ga_-\,
 T_-^{\rho'\rho}(p',p;P_0)\right] \,u_{\la}^\rho(\vp), 
\eeq
where $\ga_\pm = \half (I\pm\ga_0)$, and $T_\pm^{\rho'\rho}$ are eight
scalar amplitudes with definite parity $\pm$. 

In doing so we in particular find for the case of  the ET equations,
that the parity-conserving amplitudes $T_r^J$ satisfy
\beq
\eqlab{ETjw}
 T_r^{J\rho' \rho}=
V_r^{J\rho' \rho} +  
\frac{1}{\pi}\! 
\int\limits_0^\infty\! \dd\rq\,{\rq}^2
 \sum_{\rho''} G^{(\rho'')}_{ET}\, V_r^{J\rho' \rho''}\, 
 T_r^{J\rho'' \rho},
\eeq
where
\bea
\eqlab{GET}
G_{ET}^{(\rho)}(|\vec{q}|;P_0)&=&
2i\!\int\limits_{-\infty}^\infty\!\frac{\dd q_0}{2\pi}\,
G^{(\rho)}(\rq,q_0;P_0) \no\\
&=&   -\rho \left\{\w_q 
(-\rho P_0 + E_q + \w_q - i\eps)\right\}^{-1}.
\eea
In this work we will be focusing on solving this equation for a
one-particle-exchange \pn potential, see 
\Figref{potf}, where the potential is regulated by form factors.
In the next section we describe the $\pi N$ interaction used
in this study.

\section{Effective Lagrangian and the \pn potential}
\seclab{pot}

In the following we
specify the interaction Lagrangian of $\pi$, $\rho$, $\si$, $N$ and
isobar fields, used to construct the hadron-exchange force 
depicted in \Figref{potf} and written out in Appendix C.
The field representation and the corresponding free Lagrangian 
is chosen according to the spin and isospin of the particle.
Thus, the pion is described by scalar isovector multiplet
$\pi^a=(\pi^+,\pi^0,\pi^-)$, the sigma-meson is a scalar isoscalar field
$\si$, the $\De$(1232) is represented by a vector-spinor isoquadruplet
$\De_\mu = (\De_\mu^{++},\De_\mu^{+},\De_\mu^{0},\De_\mu^{-})$, 
and so on. 

\subsection{Nucleon and meson exchanges}
The $\pi NN$ interaction Lagrangian is taken in accordance with
the chirally-symmetric $\si$ models\cite{GeL60,Wei67}. 
In Weinberg's nonlinear realization 
the \pn scattering amplitude to the leading order   
is given by the nucleon Born term with the pseudovector coupling
plus the Weinberg-Tomozawa contact term \cite{Wei66,Wei96}. 
The pseudovector coupling reads
\beq
\lag_{\pi NN}^{{\rm (PV)}}= 
\frac{f_{\pi NN}}{m_{\pi}}\,\bar{N}\ga_\mu\ga_5
 \tau_a N\, \pa_\mu \pi^a,
\eeq
where $f_{\pi NN}$ is the pseudovector $\pi NN$ coupling constant
(the pseudoscalar coupling constant:
$g_{\pi NN}=2f_{\pi NN}\, (m_N/m_{\pi})$ will also be used below).
The Weinberg-Tomozawa contact term can be represented as 
a $\rho$-meson exchange with the following interaction:
\bea
\lag_{\rho NN}&=& g_{\rho NN}\,\bar{N}\frac{\tau_a}{2}
\left( \ga_\mu\rho^\mu_a +
\frac{i \kappa_\rho}{2m_N} \si_{\mu\nu}\, \pa^\mu \rho^\nu_a \right) N,\\
\lag_{\rho\pi\pi}&=& g_{\rho\pi\pi}\, \veps_{abc}\, 
\rho^\mu_a \pi_b\,\pa_\mu 
\pi_c,
\eea
provided the $\rho$ coupling, $g_\rho^2=g_{\rho NN}g_{\rho\pi\pi}$, 
is fixed by the KSRF relation \cite{ksrf}: 
$g_\rho= m_\rho/(\sqrt{2} f_{\pi})$, where $f_\pi \simeq 93$ MeV.
 There is also the second
form of the KSRF relation\cite{Pec68}:
$ g_\rho= \sqrt{2}\,m_\rho f_{\pi NN}/
(m_\pi g_{\rm A})$, $g_{\rm A}\simeq 1.26$,
obtained from the first one by using 
the Goldberger-Triemann relation for $f_\pi$.
It should be remarked that the Weinberg-Tomozawa contact 
term is equivalent to
the $\rho$-exchange only at threshold and
provided $g_\rho$ is fixed by KSRF relation while $\kappa_\rho=0$. 
The energy dependence is different, but not significantly
in the considered energy region.

Since we use the pseudovector $\pi NN$ coupling, the $\si$-meson is in 
principle not needed from the standpoint of chiral symmetry. 
Nevertheless, a $\si$ exchange can be used to model the isoscalar 
contribution of the correlated two-pion exchange. In order
to keep the agreement with the soft-pion theorem,
a derivative coupling to the pion is used, \ie
\bea
\lag_{\si NN}&=& g_{\si NN}\, \si\bar{N} N,\\
\eqlab{sipipi}
\lag_{\si\pi\pi}&=& -\frac{g_{\si\pi\pi}}{2m_{\pi}}\, \si\,\pa_\mu \pi_i\,
\pa^\mu \pi_i\, ,
\eea
where the sign of the $\si\pi\pi$-coupling is chosen in 
accordance with the correlated 
two-pion exchange analysis\cite{SDH94},
and is different from the one used in \cite{PeJ91}.
This interaction leads to the following on-shell potential:
\beq
\eqlab{siex1}
V_\si = \frac{g_{\si NN} g_{\si\pi\pi}}{8\pi m_\pi}
 \frac{t-2m_\pi^2}{m_\si^2-t}.
\eeq

To control the effect of the $\si$-exchange on the scattering length
we introduce a free parameter $c_\si$ in the following way:
\beq
\eqlab{siex}
V_\si = \frac{g_{\si NN} g_{\si\pi\pi}}{8\pi m_\pi} 
\frac{t-2(1-c_\si) m_\pi^2}{m_\si^2-t}.
\eeq
For $c_\si=1$  the
$\si$ contribution to the $S$-wave scattering lengths vanishes. 
Note that this modification amounts to adding the
following term to the Lagrangian:
\beq
\eqlab{primesipi}
\lag_{\si\pi\pi}'= \frac{g_{\si\pi\pi}}{2}\,c_\si m_{\pi}\si\,\pi_i^2\,.
\eeq

\subsection{$\De$-isobar exchange and higher resonances} 
The coupling of the spin-3/2 $\De$ field to the
pion and the nucleon 
is conventionally described by the following Lagrangian
 (see, e.g., Ref.~\cite{Hoh83}): 
\beq
\eqlab{conv}
\lag_{\pi N\De}= \frac{f_{\pi N\De}}{m_{\pi}}\,\bar{\De}^\mu
\left[\rg_{\mu\nu}-(z+\half)\ga_\mu\ga_\nu\right]T_a\, N\, \pa_\nu \pi^a 
+ \mbox{H.c.},
\eeq
where $z$ is the off-shell parameter, 
$T_a$ is the isospin-$\half\rightarrow\thalf$ transition operator.

As remarked in the Introduction, this coupling involves the spin-1/2
sector of the Rarita-Schwinger field, and give rise to an unphysical
spin-1/2 background. The latter effect can in principle be 
removed by inserting the spin-3/2
projection operator ``by hand'' in either the vertex or the propagator. 
For instance, Gross and Surya \cite{GrS93} have chosen this option.
However, because of the nonlocal nature of the projection operators
their use is problematic: unphysical singularities
occur at $s=0$ and $u=0$ for the $s$- and $u$-channel contribution,
respectively. In Ref.~\cite{GrS93} this problem is actually not met
because the $s=0$ point is well below the threshold, while the
$u$-graph vanishes in the approximation of that model.
If the $u$-channel $\De$ exchange is present, one usually 
prefers to keep the background and fit
the off-shell parameter \cite{PeJ91,LaA98,PaT98,kmat1,kmat2}.

We shall  also study the following 
$\pi N\De$ coupling \cite{Pas98,PT99}:
\beq
\eqlab{int3}
\lag_{\rm \pi N\De}^{{\rm (GI)}} =
 \frac{f_{\pi N\De}}{m_{\pi}m_{\De}}\,\veps^{\mu\nu\al\be}\,
(\pa_\mu\bar{\De}_\nu) \ga_5\ga_\al\,T_a\, N\, \pa_\be\pi^a 
+ \mbox{H.c.},
\eeq
referred to as {\it gauge-invariant (GI) $\pi N\De$ coupling}. 
Being invariant under the Rarita-Schwinger gauge
transformation: 
$$\De_\mu(x) \rightarrow \De_\mu (x) + \pa_\mu \eps(x),$$ 
where $\eps(x)$ is a spinor field, this coupling does not 
involve the spin-1/2 components  of
the $\De$ field. As a consequence, the spin-1/2
backgrounds are totally absent from the corresponding 
$\De$-exchange amplitudes. 

We include also the $s$- and $u$-channel graphs 
of $P_{11}$ (Roper), $S_{11}$, 
and  $D_{13}$ resonances. 
At low energies the contribution of these resonances is 
marginal,\footnote{This is generally not true for a spin-3/2 resonance 
if the conventional coupling is used, since the spin-1/2 background 
can be large even far away from the mass position.} 
but they are important for the proper description at higher energies.
The first two particles are treated same as the nucleon but
with different masses, coupling parameters, and,  in the case 
of $S_{11}$, parity. The $D_{13}$ is treated in the 
same way as the $\De$ 
(the same propagator and interaction vertex), but 
with different isospin, parity and mass. Exchanges of even higher 
spin resonances can in principle be easily included in our
model via the
amplitude obtained in Appendix D.

\subsection{Cutoff form factors}

The high-energy behavior of the hadron-exchange
field-theoretic potential is usually regulated using 
the {\it off-shell form factors} introduced in the vertices.
We have introduced them for each of the particle in  the vertex.
For the pion we use the monopole form factor:
\beq
f_\pi(k^2) = \frac{\La_\pi^2-m_\pi^2}{\La_\pi^2-k^2}.
\eeq
For the $\si$- and $\rho$-meson we use the one-boson-exchange form factor:
\beq
f_{\si,\rho}(t) = \frac{\La_{\si,\rho}^2}{\La_{\si,\rho}^2-t}.
\eeq
For the baryons we use 
the form factor of Ref.~\cite{PeJ91}:
\beq
f_B(p^2)=\left( \frac{n \La_B^4}{n\La_B^4+(p^2-m_B^2)^2} \right)^n
\eeq
with $n=2$. 

In addition, for each pion we introduce the
following cutoff:
\beq
\eqlab{regff}
f_{Regge}(\rq,s) = \frac{\La_\pi^4}{\La_\pi^4 + s \vec{q}^2} .
\eeq
This function is motivated by considering the effect
of the higher-mass states on the high-energy behavior of
the $\pi N$ propagator $G$.
If we were to include not only the pion and the nucleon but
all the states lying on the infinitely rising Regge trajectories, 
at very high
energies their effect factorizes in the form of function
like \Eqref{regff} \cite{disser}. 
The energy dependence
and the absence of singularities distinguish this cutoff from 
the usual ones, such as the monopole form. 
We take the same value $\La_\pi$ for the $ f_\pi $ and $ f_{Regge}$
cutoff masses. These two form factors do not affect
the on-shell potential, because no pion exchanges appear
in the Born graphs.

It is important to realize that the final results depend on
the off-shell form factors, even after the renormalization is applied.
The physical meaning of such form factors is usually given in analogy with
that of the electromagnetic form factors. They thus reflect the
extension of the hadrons, and in principle should be calculated
from the underlying theory.

Our fit to the $\pi N$-scattering phases 
will determine the values of the cutoff
masses. They are given in Table~\tabref{modelpar} together with the
rest of the model parameters. Using these values, in \Figref{cform1}
we have plotted the form factors which affect the loop contributions.
Their dependence on the loop momentum is shown,
while the 0-th component is fixed by the equal-time constraint and
the energy is fixed at threshold. 

The actual cutoff of the
model is given by the solid line in the figure. As one can see,
it is rather soft: it starts off as a monopole with the mass about
0.8 GeV, and is even softer above $\rq^2=0.5$  GeV$^2$. 
At higher energy it becomes softer as well, because $f_{Regge}$
is energy-dependent. However, the latter effect is small as can be
seen from \Figref{cform2}, where the energy dependence of $f_{Regge}$
 is shown
over the region of our \pn fit.
The energy dependence of $f_N(s)$ and $f_\De(s)$ is shown there as well.

\section{Renormalization}
\seclab{renorm}

Since there are s-channel singularities in the considered
potential, we have to carry out a renormalization procedure.
We adopt the scheme in which the Lagrangian is
expressed in terms of the physical parameters
and no ``bare'' parameters appear. Then,
in principle, the counter-terms should be subtracted and fixed by the
renormalization conditions. 
To perform such a renormalization procedure it is convenient to work
with the one-particle-irreducible Green functions.  One can
separate the potential into two terms $V=V_s+V_u$, where 
\beq
V_s(p',p) = \sum_{B} \Ga_B(p')\, S_B(P) \Ga_B(p)
\eeq 
represents
the $s$-channel baryon exchanges (pole terms), 
while $V_u$ contains the rest of
the graphs (nonpole terms). Since
$V_sG$ is a separable kernel, we can explicitly resum these
contributions, and find that the resulting amplitude can 
equivalently be written
as 
\beq
T(p',p) = \sum_{BB'} \Ga_{B'}^\ast(p')\, S_{BB'}^\ast(P)\, \Ga_B^\ast(p) +
T_u(p',p),
\eeq
where
\bea
\eqlab{dressvtx}
\Ga_B^\ast(p)&=& \Ga_B(p) + i\int\! \frac{\dd^4 q}{4\pi^3} \, 
\Ga_B(q) \, G(q)\, T_u(q,p),\\
\eqlab{dressprop}
(S_{BB'}^\ast)^{-1}&=&(S_B^\ast)^{-1}\de_{BB'} - \Si_{BB'}, \\
\Si_{BB'}&=&
 i\int\! \frac{\dd^4 q}{4\pi^3}\Ga_B(k) \, G(k)\, \Ga_{B'}^\ast(q),
\eea 
and $T_u$ satisfies the following integral equation,
\beq
\eqlab{bseu}
 T_u(p',p)=V_u(p',p)+ i\int\! \frac{\dd^4 q}{4\pi^3} \, 
V_u(p',q) \, G(q)\, T_u(q,p) .
\eeq

The full amplitude is thus written in terms of the
irreducible Green functions: $S^\ast, \Ga^\ast$ and $T_u$. 
The diagrammatic form of this representation is given in \Figref{irred}.

\subsection{Baryon mixing}

Note that the dressed baryon propagator, \Eqref{dressprop},
is in general nondiagonal.
In other words, the baryons can mix. Of course this mixing happens
only among the baryons with the same ``good'' quantum numbers, such as
spin and  isospin. Parity is also conserved,
nevertheless the mixing of baryons with the same spin,
isospin and opposite
parity may occur due to the negative-energy state propagation.

To perform the renormalization we need first
to diagonalize the propagator.
Since it is a complex symmetric matrix, we diagonalize it
using a complex orthogonal transformation $\cO$ ($\cO\cO^T=\cO^T\cO=$1). 
The full solution can obviously be written in the diagonal 
form as follows:
\beq
T = \sum_{B} \left(\Ga^\ast{\cO}\right)_B\,\left({\cO^T} S^\ast{\cO}\right)_B
\, \left({\cO^T}\Ga^\ast\right)_B +
T_u,
\eeq

In our model we include only two baryons with the same spin and isospin
(nucleon and $N^\ast$). For this case the
propagator is diagonalized by a 2$\times$2 complex
orthogonal matrix which can be parametrized as usual 
by one complex variable,
\beq
\eqlab{o2by2}
\cO = \left(\begin{array}{cc}
\cos \chi & \sin\chi\\
-\sin\chi & \cos \chi
\end{array}\right),
\eeq 
where in this way we introduce the $NN^\ast$ {\it mixing angle} $\chi$.

Furthermore, since we use the same Feynman rules for the nucleon 
and $N^\ast$,
their dressed vertices are equal up to the coupling constants. 
Therefore for
the $NN^\ast$ self-energy matrix one can write
\beq
\eqlab{sennstar}
\left(\begin{array}{cc}
\Si_{NN}(P) & \Si_{NN^\ast}(P)\\
\Si_{NN^\ast}(P) & \Si_{N^\ast N^\ast}(P) \end{array}\right) =
\left(\begin{array}{cc}
g_{\pi NN}^2 & g_{\pi NN}g_{\pi NN^\ast}\\
g_{\pi NN}g_{\pi NN^\ast} & g_{\pi NN^\ast}^2 \end{array}\right) \Si(P),
\eeq
while for the vertex
\beq
\eqlab{vtxnnstar}
\left(\begin{array}{c}
\Ga_{N}^\ast \\
\Ga_{N^\ast}^\ast \end{array}\right) =
\left(\begin{array}{c}
g_{\pi NN} \\
g_{\pi NN^\ast} \end{array}\right) \Ga^\ast.
\eeq
The propagator is then diagonalized by the orthogonal 
transformation \eref{o2by2}
with 
\beq
\eqlab{chinn}
\chi(P) = \frac{1}{2}\, \arctg\left\{ 
2\left(\frac{g_{\pi NN}}{g_{\pi NN^\ast}}
- \frac{g_{\pi NN^\ast}}{g_{\pi NN}}
- \frac{m_{N^\ast}-m_N}{g_{\pi NN}g_{\pi NN^\ast}\Si(P)}
\right)^{-1}\right\}.
\eeq
The corresponding eigenvalues are clearly
\bea
S_N(P)&=& \left[ \sla{P} - m_N - \Si_N(P) -i\eps \right]^{-1},\\
S_{N^\ast}(P)&=& \left[ \sla{P} - m_{N^\ast} - 
\Si_{N^\ast}(P) -i\eps \right]^{-1},
\eea
where
\bea
\Si_N&=&\left(g_{\pi NN}\cos\chi + g_{\pi NN^\ast}\sin\chi\right)^2
\Si + (m_{N^\ast}-m_N)\sin^2\chi, \no\\
\eqlab{nnstarse}
\Si_{N^\ast}&=&\left(g_{\pi NN^\ast}\cos\chi 
- g_{\pi NN}\sin\chi\right)^2
\Si - (m_{N^\ast}-m_N)\sin^2\chi.
\eea
The vertices are rotated according to
\bea
&&\Ga_{N}^\ast =
(g_{\pi NN}\cos\chi+ g_{\pi NN^\ast}\sin\chi)
\,\Ga^\ast,\no \\
\eqlab{renvmix}
&&\Ga_{N^\ast}^\ast=
 (g_{\pi NN^\ast} \cos\chi- g_{\pi NN}\sin\chi)
\,\Ga^\ast.
\eea

\subsection{Self-energy}

Let us consider the mass renormalization using the counter-term method.
The counter terms can be read off directly from the free Lagrangian.
For the spin-1/2 case, for instance, they are given by
$ Z_2 (m_0 - m) + (1-Z_2) (\sla{P} - m)$,
where $m_0$ is the bare mass, and $Z_2$ is the field
renormalization constant. The renormalized spin-1/2 baryon propagator 
is defined as 
\beq
\bS(\sla{P}) = \left[ \sla{P} - m - 
\Si^{ren} (\sla{P}) -i\veps \right]^{-1},
\eeq   
where 
$\Si^{ren} (\sla{P}) = \Si (\sla{P})-Z_2 (m_0 - m) - (1-Z_2) (\sla{P} - m)$.

In the CMS, $P=(P_0,\,\vec{0})$, the self-energy can
be written as
\beq
\Si^{ren}(P_0)=\Si_+^{ren}(P_0)\ga_+ + \Si_-^{ren}(P_0) \ga_-, 
\eeq
and  a similar decomposition holds for the propagator,
\beq
\bS(P_0) = \bS^{(+)}(P_0)\ga_+ + \bS^{(-)}(P_0) \ga_-
\eeq 
where $\ga_{\pm}=(I\pm\ga_0),$ and
\beqa
\bS^{(+)}(P_0) =\left[ P_0 - m - \Si_+^{ren} (P_0) -i\veps \right]^{-1},\\
\bS^{(-)}(P_0) =-\left[ P_0 + m + \Si_-^{ren} (P_0) +i\veps \right]^{-1}.
\eeqa
Obviously, $\ga_+$ and $\ga_-$ act as the projection operators 
onto the positive and negative energy-states, hence 
$\bS^{(+)}$ corresponds to the positive and $\bS^{(-)}$ to the
negative energy-state propagations. 

The renormalization
condition at the pole position is given by
\beqa
\left. (P_0 - m)\, \bS^{(+)}(P_0) \right|_{P_0=m}= 1,\\
\left. (P_0 + m)\, \bS^{(-)}(P_0) \right|_{P_0=-m}=  1.
\eeqa
Expanding $\Si_+(P_0)$ near 
$P_0=m$, and $\Si_-(P_0)$ near $P_0=-m,$ we find  that
 the renormalization condition requires
\bea
Z_2 (m_0 - m)&=& \Si_+ (m) =\Si_- (-m) , \no\\ 
1-Z_2&=& 
\left.\frac{\pa \Si_+ (P_0)}{\pa P_0}\right|_{P_0 = m}=
-\left.\frac{\pa \Si_- (P_0)}{\pa P_0}\right|_{P_0 = -m} .\no
\eea

As emphasized earlier \cite{PaT98},  
the above described procedure breaks down if
the self-energy is computed using a quasipotential formulation
which violates charge conjugation symmetry, since in that case
$\Si_+(P_0)\neq\Si_-(-P_0)$.
The self-energy of the spin-3/2 baryons can be renormalized similarly,
since the spin-3/2 baryon contribution to the spin-3/2
partial-waves can always be factorized
into vertices and a spin-1/2 propagator, see \Eqref{pwaj}.

\subsection{The renormalized vertex and the amplitudes}

In the adopted renormalization scheme we require 
(i) the (real part of) renormalized baryon self-energy
 $\Si^{ren}_\pm(P_0)$ and its first derivative vanish at 
the pole position $P_0=\pm m$; (ii)
the (real part of) renormalized vertex $\pi N \rightarrow$baryon 
vertex is equal to the undressed vertex at the renormalization scale
$\mu$ defined as the point
where all three particles are on the mass-shell, 
$\mu$: $k^2=m_\pi^2,\, p^2=m_N^2,\,P^2=m^2$.

 For the vertex we use the multiplicative renormalization 
 since it maintains unitarity in a simple way.
The renormalized vertex is thus defined as
\beq
\eqlab{renv}
\bGa(p;P)= Z_1 \Ga^\ast(p;P),
\eeq
where $Z_1$ is the coupling constant renormalization factor which
is readily determined from condition (ii):
\beq
Z_1=\bGa(\mu)/\Ga^\ast(\mu).
\eeq
  
In the case of the $NN^\ast$-mixing we renormalize the (scalar) function 
$\Sigma$ in Eq. ~\eref{sennstar} at the point
associated with the nucleon. This procedure clearly yields the proper 
physical 
nucleon mass pole in the corresponding baryon propagator. Adopting this
subtraction procedure the $N^\ast$ 
mass and coupling constant at the nucleon
 mass position can then be extracted.
In the various tables the values of these parameters
found in the fits are 
quoted.

After the partial-wave decomposition,
the renormalized solution of the ET equation for a given isospin $I$
and total spin $J$ and parity $r$ reads as follows
(for brevity the external momenta are omitted):
\bea
T_r^{J\rho'\rho}&=& \sum_\pi {\bGa}_r^{\rho'}\, \bS^{(r\eta_B)}\, 
\bGa_r^{\rho} \,
\de_{J_BJ}\,\de_{I_BI} + T_{u,r}^{J\rho'\rho},\\
T_{u,r}^{J\rho'\rho}& = &V_{u,r}^{J\rho'\rho} +
\frac{1}{\pi}\int\limits_0^\infty\!\dd\rq\, \rq^2 \sum_{\rho''}
V_{u,r}^{J\rho'\rho''} (\rq)\,
G^{(\rho'')}_{ET}(\rq)\,T_{u,r}^{J\rho''\rho}(\rq),\\
\eqlab{renvert}
\bGa_r^{\rho}&=& Z_1\left(\Ga_r^{\rho} + 
\frac{1}{\pi}
\int\limits_0^\infty\!\dd\rq\, \rq^2 \sum_{\rho'} \Ga_r^{\rho'}(\rq)\,
G^{(\rho')}_{ET}(\rq)\,T_{u,r}^{J\rho'\rho}(\rq) \right),\\
\eqlab{se2}
\Si_r&=& Z_1 \frac{1}{\pi}
\int\limits_0^\infty\!\dd\rq\, \rq^2 \sum_{\rho} \Ga_r^{\rho}(\rq)\,
G^{(\rho)}_{ET}(\rq)\,\bGa_{r}^{\rho}(\rq),
\eea
where $J_B,\, I_B$ and $\eta_B$ are the baryon spin, isospin and parity,
respectively. The renormalized propagator is
\beq
\bS^{(\pm)}(P_0)=\left[ \pm P_0 - m - \Si_\pm^{ren} (P_0) +i\veps \right]^{-1},
\eeq
where the renormalized self-energy is given in terms of $\Si$ of 
\Eqref{se2} [with \Eqref{nnstarse} in the case of $NN^\ast$-mixing]
 as follows:
\bea
\Si_\pm^{ren}(P_0)&=&\Si_\pm(P_0) - \Si_\pm(\pm m) \no\\
&-& (P_0 \mp m) \left.\frac{\pa \Si_\pm (P_0)}{\pa P_0}\right|_{P_0 = \pm m}.
\eea

\section{Results}
\seclab{solution}

Having described the equation for the off-shell \pn amplitudes,
its renormalization, and the driving force, we now turn to
discussing the outcome of such modeling.
For this let us first give an explicit relation 
between the on-shell 
amplitudes $T^{J\,+\,+}_r$ and the 
phase parameters.  

We introduce the standard on-shell $\pi N$
amplitudes $f_{l\pm}$, where $l=J- \half r$
is the angular momentum and $r$ is the parity of the state. 
In the normalization according to
\beq
f_{l\pm} = \frac{\eta_{l\pm} e^{2i\de_{l\pm}} -1}{2i},
\eeq
where $\de_{l\pm}$ is the phase shift and $\eta_{l\pm}$ is the
inelasticity, we can identify
\beq
f_{l\pm} = \hat{q}\,\frac{s+m_N^2-m_\pi^2}{2s}\, 
T^{J\,+\,+}_r (\hat{E},\hat{q},\hat{E},\hat{q};P_0),
\eeq 
where
\beq
\hat{q}=\sqrt{[s-(m_N-m_\pi)^2]\,[s-(m_N+m_\pi)^2]/4s},
\eeq
and $\hat{E}= \sqrt{m_N^2+\hat{q}^2}$ is the
energy of the nucleon in the CMS.

At very low energies the partial-wave amplitudes are dominated
by the threshold behavior: $\hat{q}^{2l+1}$, and their real and 
imaginary parts are related by elastic unitarity. 
Therefore, it is sometimes more useful to study \cite{Hoh95}
\beq
\eqlab{efr}
{\MM}_{lJ}(\hat{q}^2) = 
\left(\frac{m_{\pi}}{\hat{q}}\right)^{2l+1}\, \re f_{l\pm},
\eeq
instead of $f_{l\pm}$ itself or the phase shifts.
Note that, at $\hat{q}^2=0$, ${\MM}_{lJ}$ is equal to the corresponding
{\it scattering length} defined as\footnote{We shall 
commonly refer to this quantity as to scattering length, even 
though for $P$ and higher waves it is properly called scattering 
{\it volume} because of the dimension.}
\beq
a_{lJ} \equiv \lim_{\hat{q}\rightarrow 0} [\hat{q}^{-2l-1}\, 
f_{l\pm}(\hat{q})]. 
\eeq

The \pn {\it effective-range parameters}
$b_{lJ}$ can also be determined in terms of $\MM$:
\beq
b_{lJ}=  \frac{\pa}{\pa \hat{q}^2} \left. {\MM}_{lJ}(\hat{q}^2)
\right|_{\hat{q}=0}\,.
\eeq 
Instead of using this formula, we will be presenting the plot
of ${\MM}$ as a function of $\hat{q}^2$. The slope of these
``effective-range plots'' at small $\hat{q}$ indicates the
values for the effective-range parameters. In the partial waves
that support a resonance (e.g., $P_{11}$, $P_{33}$) it is
more appropriate to study another effective-range expansion:
\beq
\hat{q}^{2l+1} \cot \de_{l\pm}= \frac{1}{a_{lJ}} +\frac{1}{2}
r_{lJ}\, \hat{q}^2 +\cdots,
\eeq
nontheless, we only will address threshold parameters $a$ and
$b$ for all partial waves.

\subsection{The $K$-matrix approximation}

In this subsection we focus on the {\it K-matrix approximation}
to the full scattering problem. Results of the full model are
discussed in the next subsection. 
In the K-matrix approximation $\pi N$  amplitude 
is given by the 
lowest order $K$-matrix expression:
\beq
\eqlab{kmatrix}
f_{l\pm} = \frac{K_{l\pm}}{1-i K_{l\pm}},
\eeq
where
$ K_{l\pm} = \hat{q}\, \al(s)\, V_\pm^J, \,\, l=J\mp \half$,
and $V_r^J$ is the partial-wave potential [obtained from the 
potential by means of Eqs.~\eref{defparity} and \eref{f1}].
Equation \eref{kmatrix} clearly satisfies elastic unitarity, 
but the principal
value of the loop integrals is neglected.  

This approximation is considered to be good, at least at low energy,
and has been frequently used, see e.g.\ \cite{kmat1,kmat2} for
most recent applications to the \pn scattering. At low
energies, indeed, the soft-pion theorems dictate that the Born 
graphs dominate, implying  for the potential modelling that the
rescattering effects should be relatively small. When the latter is
true, considering the K-matrix approximation may allow us
to make a preliminary adjustment of some model parameters
without going to the full calculations. We in particular
would like to examine the effect of using different
$\pi N\De$ couplings in the $\De$ exchange contribution.

As is well known,
the S-wave scattering lengths are well reproduced by
the Born-level nucleon and $\rho$-meson exchanges 
alone \cite{Wei66,Wei96}. Taking $g_{\pi NN}^2/4\pi =13.6$
and $g_{\rho}^2/4\pi = 3.0$,
we plot in \Figref{plot0} the contribution of these three graphs
($s$- plus $u$-channel nucleon exchange plus $t$-channel $\rho$ exchange)
to $\MM_{lJ}$ for all the $S$ and $P$ partial waves. 
The figure shows that the $S$-wave scattering lengths 
are indeed reproduced. 
However, the energy dependence of the $S$ waves is not well described.
In fact, the slopes
of the effective-range plots have
a wrong sign. Furthermore, the $P$ waves are not reproduced at all. 
Similar
picture occurs if the contact term is used instead of
the $\rho$ exchange. 

Looking at the ratio
of the $S$-wave lengths it is certainly plausible that they
should be dominated by some isovector contribution,\footnote{
Experimentally the ratio $a_{S11}/a_{S31}\approx -1.75$, which is
close to the ratio of the isospin factors for an isovector meson
contribution.} such as the $\rho$-meson exchange. Therefore, 
it would be interesting to
find a simple mechanism which accounts for both the $P$ waves and the
energy behavior of the $S$ waves, and, at the same time, does not
affect the $S$-wave scattering length. Since  $P_{33}$ has the largest
discrepancy, we study first the effect of the 
$\De$-isobar exchange.

In \Figref{plot1} we show the calculations performed 
with the two different choices of the off-shell parameter: 
$$\begin{array}{ll}
 z=-\quarter & \mbox{(Peccei choice \cite{Pec68})},\\
z=\half &\mbox{(NEK choice \cite{Nat71})},\end{array}
$$
and $f_{\pi N\De}^2/4\pi =0.36$.
The nucleon and $\rho$ coupling constants are kept the same as
in the previous calculation, with $\kappa_\rho=3.7$. 

One can see that, as far as $P_{33}$ is concerned, 
the $\De$ contribution is very 
plausible for both choices of $z$.
The large difference between the choices
clearly shows up in the spin-1/2 partial waves, where the
$\De$ exchange produces a significant background contribution
controlled by the off-shell parameter. 

In particular,  Peccei's choice affects 
substantially the $S$ waves, and hence  spoils the scenario where
those are dominated by the $\rho$-exchange. The $P$ waves look 
much better,
which could be due to the remarkable fact that, for this
particular choice, the  $s$-channel $\De$-exchange
graph gives only a tiny contribution to the
spin-1/2 $P$ waves. 
The NEK choice, in contrast, does not affect
the $S$ waves at threshold and gives a large 
effect in the $P$ waves. 

Clearly, Peccei's choice could be favorable 
phenomenologically as long as the missing strength in the $S$ waves
is somehow explained; for instance, by an isoscalar meson exchange. We
believe such a scenario is realized in most of the
models which use coupling \eref{conv} and describe the scattering
lengths correctly. However, apparently 
it is not possible to describe simultaneously
the $S$- and $P$-wave scattering lengths 
in the tree-level model with only
the $\rho$, $N$ and $\De$ exchanges.

The GI $\pi N\De$ coupling \Eqref{int3}, in combination
with the usual Rarita-Schwinger propagator, leads to the
$\De$-exchange amplitude with only spin-3/2 contributions
(omitting the isospin-dependent factor):
\bea
\eqlab{giample}
&&V_{\De,\,s{\rm -exch}}=\frac{f_{\pi N\De}^2}{4\pi m_{\pi}^2}
\left(\frac{s}{m_\De^2}\right)\frac{\sla{P} + m_\De}{s-m_\De^2} \, 
{\mathscr P}^{3/2}_{\al\be}(P)\, {k'}^\al k^\be ,\\
&&V_{\De,\,u{\rm -exch}}=\frac{f_{\pi N\De}^2}{4\pi m_{\pi}^2}
\left(\frac{u}{m_\De^2}\right)\frac{\slap_u + m_\De}{u-m_\De^2}
\, {\mathscr P}^{3/2}_{\al\be}(p_u)\, 
{k}^\al {k'}^\be ,
\eea
where $p_u= P-k-k'$, and 
$$ {\mathscr P}^{3/2}_{\al\be}=\rg_{\alpha\beta} - 
\third\gamma_{\alpha}\gamma_{\beta}
       - \frac{1}{3p^2}(\slap\gamma_\alpha p_\beta + 
p_\alpha\gamma_\beta \slap) $$ 
is the spin-3/2 projection operator.

A calculation using this amplitude is shown in \Figref{plot1},
in comparison with analogous calculations using the
conventional coupling with Peccei and NEK choices of the
off-shell parameter. First of all we remark that the 
$\De$ contribution to the spin-1/2 partial-waves comes 
from the $u$-graph only, and not from the spin-1/2 components.
From the figure we can see that the GI and NEK coupling
produce similar contributions to the $S$-waves, but largely
different results in the spin-1/2 $P$-waves. 
The resonant
$P_{33}$ wave comes out very much alike for both GI and conventional
couplings, the main difference being $(s/m_\De^2)$ factor
in front of the GI amplitude. Hence, at threshold the
GI result is factor of $(m_N+m_\pi)^2/m_\De^2\simeq 0.76$
smaller than the conventional result
(modulo small contributions from the $u$-channel graph, which
for instance are responsible for the difference between the 
Peccei and NEK choice in the $P_{33}$ wave). Despite that,
after a readjustment of parameters the GI invariant coupling
usually gives a better description of the scattering lengths,
see calculations presented by Tables~\tabref{sl1} and \tabref{cc1}
(note that here the NEK choice has been used with $\kappa_\rho=0$,
as is suggested in the original paper \cite{Nat71} and indeed gives
a better fit than with $\kappa_\rho=3.7$).
The problem with the wrong energy behavior of the $S$ waves, however,
applies to all these calculations. To correct for this a scalar
$\si$-meson exchange is needed in  the $\pi N$ force.

Inclusion of the $\si$-exchange allows us to fit the
scattering lengths to practically arbitrary accuracy,
independently of whether we use the conventional (model A) 
or the GI coupling (model B), see Tables~\tabref{sl2} and \tabref{cc2}.
Since we have fixed $c_\si=1$, and thus do not allow
the $\si$ to affect the $S$-waves, the best fit of the
off-shell parameter give the NEK value which also
has vanishing $S$-wave contribution. The $S$-wave lengths
are therefore explained in both models A and B in exactly the
same way: by $\rho$ and nucleon exchanges alone. The small
difference between the NEK and GI coupling in $P_{33}$
(see \Figref{plot1}) is apparently compensated by
the difference in $f_{\pi N\De}$. The large differences
in the other three partial-waves has mostly been removed by taking
a different value for $\kappa_\rho$. 
This example shows that if the potential is general enough, 
the spin-1/2 backgrounds of the $\De$ can possibly be reshuffled into 
other contributions.

\subsection{Full calculations}
We solve the ET equation, \Eqref{ETjw}, by Pad\'e approximants 
following the procedure described in  Refs. \cite{NiT68,FlT75}. 
Writing the equation as
\beq
T=V+VGT,
\eeq
we begin by performing several iterations of the potential, and hence
find the first few terms in the  expansion of the amplitude:
\beq
T= \sum_{n=0}^{\infty} T^{(n)} = V+\sum_{n=1}^{\infty} V(GV)^n . 
\eeq  
The solution is then sought in the form of the Pad\'e approximant.
For the equal-time equation with the model
potential the solution accurately converges by performing just six
iterations.
 
We have fitted the on-shell solution to
the KH80 \cite{KH80} and SM95 \cite{ASW95} 
\pn scattering partial-wave analyses, in the region from the
threshold up to 600 MeV pion kinetic energy in the lab.
The resulting fit is shown by the solid lines in \Figref{pin600f}.
The determined coupling constants and masses
are given in Table~\tabref{modelpar}. 

The dashed line in the $S_{11}$ phase-shift of \Figref{pin600f}
indicates the calculation
without the $s$-channel $S_{11}$ resonance graph. This graph contributes
also to the $P_{11}$ wave but calculation with or without it produce
practically identical results. The $S_{11}$ resonance pole is thus 
relevant only for the $S_{11}$ wave above 400 MeV.

Up to 350 MeV the agreement of the model with the partial-wave data
is very good as can be seen in \Figref{pin350f}.
At energies exceeding 350 MeV inelastic channels become important. 
Since we have not considered any inelastic mechanisms, some 
discrepancies seen in \Figref{pin600f} at higher energies 
are not surprising.

In fitting we have paid careful attention to the low-energy
behavior, particularly to the correct shape of the 
``effective-range plot'',
${\MM}_{l\pm} (\hq^2)$, defined in \Eqref{efr}. The model description
of these plots is shown in \Figref{effr}. The scattering lengths
can be read off these plots at $\hq=0$. The $\chi$-square value 
for the scattering lengths with respect to the SM95 analysis is $1.4$.

To give a feeling about the size of the rescatterings in the model, 
the dashed lines in the figures indicate
the results of the calculation where the principal part of the 
rescattering integrals is neglected, \ie the $K$-matrix approximation.
Unlike in the $K$-matrix calculations of the previous subsection,
here the form factors are included, and the same 
set of parameters is used as in the full calculation. 

The large difference between the
full and the $K$-matrix calculation
in the waves where the baryon pole contributions are present indicates,
therefore, that the ``dressing'' and renormalization 
of the pole contributions has
a significant effect. The effect of the rescatterings in the
nonresonant waves, such as $S_{31}$, $P_{31}$, etc., which have
contribution only from the $T_u$ term, is smaller.
We also  observed 
that the ``attractive waves'' (which have positive scattering length) 
receive comparably large and positive
rescattering contributions at threshold.

For the resonant waves the $T_u$ contribution may lead to significant
shifts of the resonance position. For instance, the $\De$-pole 
in the P$_{33}$ is located at $\sqrt{s}=m_\De=1.252$ GeV, while
inclusion of $T_u$ leads to the physical $P_{33}$ wave which resonates
at $\sqrt{s}=1.232$ GeV. In the $K$-matrix approximation the resonance
always occurs at the pole position. To describe $P_{33}$
in this approximation one is to use $m_\De=$1.232 GeV.

By readjusting the parameters we are able to reproduce  the 
phase-shifts in the $K$-matrix approximation up to 350 MeV, 
see \Figref{kmat350f}. 
The values of the parameters are given in Table~\tabref{kmatpar}.
Note, however, that using the full model we could achieve 
the fit of a better
quality and up to higher energies.

The ($s$-channel) baryon pole terms are modified
by the vertex and self-energy
corrections. As has been seen from comparing the full and $K$-matrix
calculation,  such a ``dressing'' may have
appreciable effects, especially in the resonance
$P_{11}$ and $P_{33}$ waves. 
In \Figref{sef} we plot the real and imaginary
part of the nucleon and the $\De$ isobar self-energies. From the figure we
can see that the energy dependence is indeed significant. The same is
observed for the $N N^\ast$ mixing angle plotted in \Figref{nnmixf}.
The renormalization of the pole terms produces the values of the
renormalization constants given in Table~\tabref{renpar}.

The vertex corrections are studied using the dynamical form factor
defined as follows:
\beq
\eqlab{dynffs}
F^\rho (\rq^2,s)={\bf \Gamma}^{\rho}(\rq^2,s)/\Ga^{\rho}(\rq^2,s),
\eeq
where $\Ga^{\rho}$ is the undressed off-shell vertex,
and ${\bGa}^{\rho}$ is the renormalized off-shell vertex, see 
\Eqref{renvert}.
Note that
the coupling constants and the cutoff form factors 
are cancelling out in the expression \eref{dynffs},
since they are the same for both of the vertices.
The dynamical form factors are thus equal to unity at the 
renormalization point.

The model prediction for  $\pi NN$, 
$\pi N\De$ and  $\pi ND_{13}$ 
form factors  is given
in \Figref{form1} as a function of the off-shell 3-momentum $\vec{q}^2$
for $\sqrt{s}=m_N+m_\pi$, and in \Figref{form2} as the function
of energy for the on-shell situation, $|\vec{q}|=\hq$. According to these
figures the rescatterings have much larger effect on the energy
dependence then on the off-shell behavior of the $\pi N$ state.  

Recently, the  $\pi NN$ form factor has been studied
by Saito and Afnan \cite{SaA94}, and by Sch\"ultz, Haidenbauer 
and Holinde \cite{SHH96} in a similar modelling. In the latter work  
the $\pi N\De$ form factor has been studied as well. 
To compare with the results presented in \cite{SHH96}, we need
to multiply our dynamical form factor by the off-shell form factor
given in \Figref{cform1} (solid line). We then can see that
the resulting $\pi NN$ form factor of our model agrees in the 
main features with that of model \cite{SHH96}.  It is therefore
less soft than the form factor found in Ref.~\cite{SaA94}.
This allows the rescattering contributions to play a bigger 
role.

\section{Discussion and conclusion}
\seclab{concl}

Throughout the calculation we have been fixing the $\pi NN$
coupling constant to the value advised by the
Nijmegen group \cite{STS93}: $f_{\pi NN}^2/4\pi = 0.0757$.
Our fits were not very sensitive
to the increase of the coupling towards more traditional
value of 0.078.

The value of $g_\rho$ comes out close to the one 
 inferred by the $\rho$-meson decay width: 
$g_\rho^2/4\pi \simeq 2.8$. It is also consistent
with the KSFR relation (Sec. IV A), which gives 2.78 in its first form,
and 2.94 in the second form. The small value of $\kappa_\rho$
has been mainly dictated by the simultaneous fit to $S_{31}$ and
$P_{31}$ waves at the higher energy scale. At low energies the change in
$\kappa_\rho$ affects mostly the $P_{11}$ channel, as can be seen
from  \Figref{effr} where the dash-dotted line
represents the calculation with the vector meson dominance
value: $\kappa_\rho=3.7$.

Comparing Tables~\tabref{modelpar}  
and \tabref{kmatpar} we see that, depending on whether
the rescattering  contributions are included or not, very different
values of the $\De$-isobar masses and coupling strengths
are needed to obtain correct phase shifts. This indicates
how the dynamical component due to \pn loops may play
a significant role in the generation of the observed
$\De$(1232) resonance. That is
in addition to the ``elementary'' component due to
the formation of the three-quark state.

In comparing with other 
relativistic models we can comment that our
major difference with the model of Pearce and Jennings \cite{PeJ91} 
resides in the $\De$-isobar
and $\si$-meson contributions. For the $\De$, they 
use the conventional coupling, and for the $\si\pi\pi$ coupling
they use \Eqref{sipipi} with the opposite sign and without the
additional $c_\si$ term \Eqref{primesipi}. The $\si$ contribution is thus
attractive in their case and can account for the discrepancy in 
the $S$-waves
which appears due to the $\De$ background. 
Although it probably gives rise
to problems at higher energies, which is indicated by the very soft
cutoff form factor for the $\si$, with cutoff mass $\La_\si = 0.5$ GeV,
used in \cite{PeJ91}. 

Gross and Surya \cite{GrS93} made
a static approximation of all the $t$- and $u$- channel graphs. 
Their approximation leads to separability of the
potential and hence the complexity of solving
the integral equation numerically is avoided. It also simplifies 
considerably
the subsequent photoproduction analysis, since the meson- and 
isobar-exchange
currents have a form of contact terms.
On the other hand,  the spin-3/2 (and higher) waves, such as $P_{33}$
and $D_{13}$, receive  contributions only from the
$s$-channel exchanges of corresponding resonances, which
in particular leads to overestimates of the resonance 
coupling parameters.
Also, the static (zero-range) approximation of 
the $u$-exchange potential can
be justified only at low energy, because in principle 
the range of such a potential
rapidly increases from $1/\sqrt{2 m m_\pi}$ (where $m$ is the 
exchanged particle mass) at threshold till 
$1/m_\pi$ at high energy. We observed that already at 100 MeV 
pion kinetic
energy the solution of the integral equation for the static or
exact $u$-exchange potential differ significantly.
In contrast, the range of the $t$-exchange
potential is determined only by the exchange mass, and if that
mass is heavy enough the static approximation may be applicable.
Another difference comes from the fact that Gross and Surya
use the admixture of pseudoscalar and pseudovector coupling
for the $\pi NN$ vertex. Consequent  
differences, motivated by the consistency with the soft-pion
limit, appear in the form of the $\rho$ and $\si$ couplings.  

Lahiff and Afnan \cite{LaA98} do not include resonances beyond
the $\De$(1232) but apart from that they use an interaction,
which is very similar to ours.  They have as well 
compared the conventional versus gauge-invariant 
$\pi N\De$ coupling. However, 
in contrast to us, they preferred the former one, particularly
because the spin-1/2 background adds a significant attraction
in the $P_{11}$ channel, which helps to simulate partially 
the Roper-resonance behavior. We include this resonance explicitly 
and hence the spin-1/2 background is not necessary for fitting
$P_{11}$ phase shift.

Including the $\De$ and $D_{13}$ resonance
in the Lagrangian via the relativistic spin-3/2 fields,
we have used couplings which do not involve the spin-1/2
sector of the Rarita-Schwinger theory, and therefore
no ``spin-1/2 backgrounds'' associated with
these particles appear in the $S$-matrix. We do not
need these backgrounds to obtain a proper description of
the data. On a simple example (see Tables \tabref{sl2} and
\tabref{cc2}) we have seen that, even when
these backgrounds may sometimes seem relevant for the description,
their role can be taken by other mechanisms, hopefully with more sensible
physical interpretation. 

 Although the presented model improves in some aspects
previous relativistic analyses based on the potential approach, 
the difficulty of controlling the chiral symmetry 
constraint is present here as well. We do begin with a
Lagrangian, and thus the driving force, consistent with 
chiral symmetry, but this consistency can in principle be 
spoiled by rescatterings, particularly because of 
the loss of crossing symmetry. However, numerical checks
(see, e.g., Ref.~\cite{PeJ91})
indicate that the amount of violation of the soft-pion theorems
is usually negligible in the potential modeling. Especially
if to take into account that the prime objection of
such models is to describe the \pn physics at intermediate
energies where unitarity aspects take the leading role.  
It would, of course, be anyhow important to build
the chiral constraint more precisely into the nonperturbative
\pn models.

In conclusion, we have obtained a description of the $\pi N$
force in a relativistic dynamical model based on the covariant
equal-time (quasipotential) equation for a
hadron-exchange potential. 
The good quality of our fits in the region up to 600 MeV pion 
kinetic energy suggests the used force and the relativistic 
approach may be 
considered  reasonable, even though this model still lacks inelastic
mechanisms which can become important in $S_{11}$ and $P_{11}$ channels
above 400 MeV. It is clearly of interest to study 
the model predictions in other processes, such as
pion photoproduction and Compton scattering in the $\pi N$ system.

\acknowledgments

This work was partially financially supported by de Stichting voor
Fundamenteel Onderzoek der Materie (FOM), which is sponsored by 
Nederlandse Organisatie voor Wetenschappelijk Onderzoek (NWO).

\appendix

\section{Conventions}
Metric 
$\rg_{\mu\nu} = {\rm diag}(+1,-1,-1,-1)$;  
Levy-Cevita symbols: $\veps^{0123}=1$, $\veps^{0ijk}=\veps^{ijk}$.

Pauli spinors: 
in the direction $\theta,\,\vfi$ are given by
$$
\chi_{\la}(\theta,\vfi) = \sum_{\la'=-1/2}^{1/2} d^{1/2}_{\la\la'}
(\theta)\,e^{i(\la-\la')\vfi}\, \chi_{\la'}(0),
$$
where $\chi^\dagger_{1/2}(0)= (1,0)$, 
$\chi^\dagger_{-1/2}(0)= (0,1)$,
and $d^J_{\la\la'}(\th)$ are the Wigner $d$-functions.

Positive- and negative-energy helicity spinors:
\bea
  u_{\la}^{(+)}(\vec{p}) & = & \left[ \begin{array}{c} N_+ \\ 
                         2\la N_-  \end{array} \right]
              \otimes  \chi_{\la} (\theta,\vfi) , \nonumber \\
  u_{\la}^{(-)}(\vec{p}) & = & \left[ \begin{array}{c} 
                           - 2\la N_- \\
                            N_+ \end{array} \right]
          \otimes      \chi_{\la}(\theta,\vfi), \no
\eea
where $N_\pm = \sqrt{(E_p\pm m)/2 E_p}$, $E_p=\sqrt{m^2+\rp^2}$, 
and $\rp$, $\theta$, $\vfi$ are the spherical coordinates of
$\vec{p}$. 
The helicity spinors satisfy the following
orthogonality and completeness conditions:
\bea
&& u_\la^{\rho\dagger} ( \vec{p})\, u_{\la '}^{\rho'} (\vec{p})
= \delta_{\rho \rho '} \delta_{\la \la '}, \no\\
&&\sum_{\rho=\pm}\sum_{\la} u_\la^{\rho} ( \vec{p})\, 
u_{\la}^{\rho\dagger} (\vec{p}) = 1. \no
\eea

\section{Partial-wave off-shell \pn amplitudes}

The partial-wave reduction is done in the CMS, 
where the total four-momentum $P=(P_0,\vec{0})$.
Using the orthonormality and completeness of the 
nucleon helicity spinors (see Appendix A),
we write the BS equation for the {\em helicity amplitudes}:
\beq
\eqlab{BSEh}
 T_{\la' \la}^{\rho' \rho}(p',p)=
V_{\la' \la}^{\rho' \rho}(p',p) +  i\sum_{\rho'' \la''} 
\int \frac{\dd^4 q}{4\pi^3}
 \, V_{\la' \la''}^{\rho' \rho''}(p',q) \, G^{(\rho'')}(q)\,
 T_{\la'' \la}^{\rho'' \rho}(q,p),
\eeq
where we have assumed that 
the \pn propagator is diagonal in the helicity basis, \ie
\beq
\bar{u}^{ \rho'}_{\la'}(\vq) \gamma_0 G(q) \gamma_0
 u^{\rho''}_{\la''}(\vq) = \delta_{\la' \la''}\,
\delta_{\rho' \rho''}\, G^{(\rho'')}(q),
\eeq
which is true in the CMS.
\Eqref{BSEh} yields (omitting the momenta) 
\beq
\eqlab{BSEpwh}
 T_{\la' \la}^{J\rho' \rho}=
V_{\la' \la}^{J\rho' \rho} +  
\frac{i}{\pi^2}\! 
\int\limits_{-\infty}^\infty\!\dd q_0
\int\limits_0^\infty\! \dd\rq\,{\rq}^2\,
\sum_{\rho'' \la''} G^{(\rho'')}\, V_{\la' \la''}^{J\rho' \rho''}\, 
 T_{\la'' \la}^{J\rho'' \rho},
\eeq
where 
\bea
\eqlab{pwd1}
&&X_{\la'\la}^{\rho'\rho}(p',p;P_0) = \sum_{J=1/2}^\infty (J+\half)\,
X_{\la'\la}^{J\rho'\rho}({p}_0',\rp',p_0,\rp;P_0)\, d^J_{\la'\la}(\th),\\
&&\eqlab{pwd2}
X_{\la'\la}^{J\rho'\rho}({p}_0',\rp',p_0,\rp;P_0)=
\int_{-1}^{1}\!\dd(\cos \th)
\,X_{\la'\la}^{\rho'\rho}(p',p;P_0)\, d^J_{\la'\la}(\th),
\eea
with $X=V$ or $T$.
We have chosen the 3-vectors $\vp$ and $\vp'$ to lie in the $XZ$-plane
(hence $\vfi=\vfi'=0$), and $\th$ is the center-of-mass scattering angle.

{\it Parity conservation} infers 
the following symmetry for the partial-wave helicity amplitudes:
\beq
\eqlab{pconserve}
 T_{\la'\la}^{J\rho'\rho}(p_0',\rp', p_0,\rp;P_0) = 
\rho'\rho\,T_{-\la',\,-\la}^{J\rho'\rho}(p_0',\rp', p_0,\rp;P_0).
\eeq
These relations reduce the number of independent amplitudes 
from  sixteen to eight. 
{\it Time-reversal invariance} implies 
\beq
 T_{\la'\la}^{J\rho'\rho}(p_0',\rp', p_0,\rp;P_0) = 4\la'\la\, 
T_{\la\la'}^{J\rho'\rho}(p_0',\rp', p_0,\rp;P_0), 
\eeq
which obviously does not give any new relations.

It is convenient to 
introduce the partial-wave state with definite parity $r$ 
in terms of the partial-wave helicity state \cite{JaW59,Kub72}: 
\beq
\eqlab{parstate}
\left. |J,r,\rho\right> = 
\frac{\left. |J,\rho,\la\right> + r\rho 
\left. |J,\rho,-\la \right>}{\sqrt{2}} .
\eeq
The BS equation for the parity-conserving amplitudes takes the form
\beq
\eqlab{BSEjw}
 T_r^{J\rho' \rho}=
V_r^{J\rho' \rho} +  
\frac{i}{\pi^2}\! 
\int\limits_{-\infty}^\infty\!\dd q_0
\int\limits_0^\infty\! \dd\rq\,{\rq}^2\,
\sum_{\rho''} G^{(\rho'')}\, V_r^{J\rho' \rho''}\, 
 T_r^{J\rho'' \rho},
\eeq
where $r$ denotes parity.
Note the simplification: \Eqref{BSEjw} has two less coupled channels
than \Eqref{BSEpwh}. 
The partial-wave decomposition of the ET equation \eref{compact}
proceeds in the same way. In addition we use in this case the fact that
in the CMS $V$ and $T$ are independent of $q_0$.  The resulting
equations are given by \Eqref{ETjw}.

The amplitudes \eref{defparity} are  related in 
a simple way to the parity-conserving 
partial-wave amplitudes of Jacob and Wick \cite{JaW59}. 
Starting from representation \Eqref{defparity}, we first reduce
the spinor algebra to the subspace of Pauli spinors $\chi_\la(\th)$
and use
\beq
\chi^\dagger_{\la'}(\th')\, \chi_\la(\th) = d^{1/2}_{\la'\la}(\th'-\th).
\eeq
With the aid of Eqs.~\eref{pwd2}, \eref{parstate} and the identity
\beq
d^{1/2}_{\la'\la}(\th)\,d^{J}_{\la'\la}(\th) =
\frac{1}{2}\left( P_{J-\half}(\cos\th) + 4\la\la'\,P_{J+\half}(\cos\th)\right),
\eeq
we find the sought expression for the partial-wave 
amplitude $T_{r}^{J\rho'\rho}$ with definite parity $r$ 
in terms of $T_\pm$: 
\beq
\eqlab{f1}
T_r^{J\rho'\rho} =
\frac{1}{2}\int_{-1}^1 \!\dd (\cos\th)\,\left[ T_r^{\rho'\rho}(\th)
\,P_{J-\half}(\cos\th)\, R^{\rho'\rho}_{r} +
T_{-r}^{\rho'\rho}(\th)
\,P_{J+\half}(\cos\th)\, R^{\rho'\rho}_{-r}\right],
\eeq
where the dependence on external momenta is omitted
for brevity and only the dependence
on the scattering angle is exhibited. The factors
\bea
\eqlab{rfactors}
&& R^{\rho'\rho}_{r} = \left\{\begin{array}{lc} 
 \rho\rho'N_\rho N_{\rho'}',& r=+\\
- N_{-\rho} N_{-\rho'}',&r=- \end{array}\right. \\
&& N_\pm = \frac{E_p\pm m_\rN}{2E_p},\,\,\,\,\,
N_\pm' = \frac{E_{p'}\pm m_\rN}{2E_{p'}},
\eea 
arise from the helicity spinors.

Now we can write down the relations due to 
the {\it charge conjugation symmetry}.
It relates the amplitudes of positive energy with those of
negative energy and opposite parity: 
\beq
T_r^{J(\rho',\rho)}({p}_0',\rp',p_0,\rp;P_0)=
T_{-r}^{J(-\rho',-\rho)}(-{p}_0',\rp',-p_0,\rp;-P_0).
\eeq
Needless to say, these relations will only hold in the
quasipotential formulations which satisfy the charge 
conjugation symmetry.

The {\it isospin decomposition} of the $\pi N$ amplitude
is carried out as follows.
If we denote $\chi_N$ and $\phi_\pi^a$ as the
isospin states of the nucleon and pion, respectively,
then
\bea
T&=& {\phi'}_\pi^a 
\chi_N'  \left[ \de_{ab}\, T^{(+)} + i\veps_{abc}\tau_c\,T^{(-)}
\right] \chi_N \phi_{\pi}^b \no\\
&=& {\phi'}_{\pi}^a \chi_N'  \left[\third \tau_a\tau_b \,
 T^{1/2} + 
(\de_{ab}- \third \tau_a\tau_b)\, T^{3/2}
\right] \chi_N \phi_{\pi}^b, 
\eea
where $\tau_a$ ($a=1,2,3$) are the isospin Pauli matrices, satisfying
$\half [\tau_a, \tau_b]= i\veps_{abc}\tau_c.$ Evidently,
\beq
T^{(+)}=\third (T^{1/2}+2 T^{3/2}),\,\,\,\,
T^{(-)}=\third (T^{1/2}-T^{3/2}).
\eeq

\section{The off-shell $\pi N$ potential}
According to \Eqref{offt}
the off-shell $\pi N$ amplitude can completely be specified by
the scalar $2\times 2$ matrices $A$ and $B$. 
Here we give the expressions for these matrices corresponding to
the tree-diagram potential in \Figref{potf}. Also the isospin
factors, ${\cal F}(I)$, are given.

\subsection{Baryon-exchange graphs}
\ni
For the isospin-1/2 baryon: ${\cal F}(\half)=-1$, ${\cal F}(\thalf)=2$.

\ni
For the isospin-3/2 baryon: ${\cal F}(\half)=2$, ${\cal F}(\thalf)=\half$.

\ni
{(a)}  The $u$-channel exchange of a baryon with spin $\half$, 
mass $m$ and
parity $\eta$, using the (pseudo-)scalar vs (pseudo-)vector
admixture coupling, cf.\ Ref.\ \cite{GrS93},
specified by parameter $\la$ [$\la=0$ corresponds to pure
(pseudo-)vector and is used in the text]:
\bea
A^I&=&\frac{g_{\pi NB}^2}{4\pi}\frac{1}{u-m^2}\left[
\la\bmat  m & \eta \\ 
       \eta & 0 \emat  
+ \frac{\la(1-\la)}{2m} \bmat -2\eta u +\eta(p^2+{p'}^2)&-m \\
-m&2\eta\emat \right.\no\\
&&+\left. \frac{(1-\la)^2}{4m^2}
\bmat m(u-p^2-{p'}^2)& \eta(p^2+{p'}^2)\\
\eta(p^2+{p'}^2)&-m \emat\right] {\cal F}(I),\\
B^I&=&\frac{g_{\pi NB}^2}{4\pi}\frac{1}{u-m^2}\left[
\la\bmat  -\eta & 0 \\ 
       0 & 0 \emat  
+ \frac{\la(1-\la)}{2m} \bmat 2m &-\eta \\
-\eta&0\emat \right.\no\\
&&+\left. \frac{(1-\la)^2}{4m^2}
\bmat -\eta u& m\\
m & 1 \emat\right] {\cal F}(I).
\eea

\ni
{(b)} The $\De$ exchange using the conventional coupling,
\Eqref{conv}:
\bea
A^I&=&\frac{f_{\pi N\De}^2}{4\pi m_\pi}\frac{1}{u-m^2}\left\{
A_{P\thalf} \right.\nonumber\\
&+& \frac{2}{3m^2}\left[ a^2 
\bmat 0& p'\cdot p_u-\bar{P}_{22}-{p'}^2\\
p\cdot p_u-\bar{P}_{22}-{p}^2&0\emat \right.  \no\\
&-& 2azm\bmat \bar{P}_{22}-({p'}^2 p+p^2p')\cdot p_u/u&0\\
0&1-(p+p')\cdot p_u/u\emat\no\\
&+&m(\half-2az) \bmat ({p'}^2 k+p^2k')\cdot p_u/u
-(k+ k')\cdot p_u-2 P_{22}&0\\
0&(k+k')\cdot p_u/u\emat  \no\\
&+&  a(1+a)\bmat 0& k'\cdot p_u+2P_{22}\\
k\cdot p_u+2P_{22}&0\emat  \no\\
&-& \left.\left. (1+a)^2 P_{22}\bmat 0&1\\
1&0\emat - m(1+2az) P_{22} \bmat 1 &0\\
0&0\emat
\right]\right\} {\cal F}(I), \\
B^I&=&\frac{f_{\pi N\De}^2}{4\pi m_\pi}\frac{1}{u-m^2}\left\{
B_{P\thalf} \right.\nonumber\\
&+& \frac{2}{3m^2}\left[ a^2 
\bmat \bar{P}_{22}& 0\\
0&1\emat   -2azm\bmat 0& p'\cdot p_u/u\\
p\cdot p_u/u&0\emat \right. \no\\
&+&m(\half-2az) \bmat 0&-k\cdot p_u/u\\
k'\cdot p_u/u&0\emat  - a(1+a)\bmat  (k'+k)\cdot p_u+2P_{22}&0\\
0&0\emat  \no\\
&+&\left.\left.  
  (1+a)^2 P_{22}\bmat 1&0\\
0&0\emat 
\right]\right\} {\cal F}(I),
\eea
where
\bea
&& a=-z-\half ,\no\\
&&P_{22}=(k'\cdot p_u)(k\cdot p_u)/u,\,\,\,\,
\bar{P}_{22}=(p'\cdot p_u)(p\cdot p_u)/u,\no\\
&& p_u=p-k'=p'-k,\,\,\,\, u=p_u^2,\no
\eea
and $A_{P\thalf}$, $B_{P\thalf}$ are the contributions of 
the spin-3/2 projection operator. 

For half-integer spin
the contributions of the spin-$j$ projection operator read
\bea
A_{Pj}&=&\frac{(-1)^{j-\half}(j-\half)!}{(2j)!!}
(\tk^2 \tilde{k'}^2)^{(j-\half)/2} \,\left[\bmat m&1\\
1&0 \emat P_{j+\half}'(x_u) \right.\no\\
&&\left. +\bmat m[\bar{P}_{22}-({p'}^2 p+p^2p')\cdot p_u/u]&\,\,
p'\cdot p_u-\bar{P}_{22}-{p'}^2 \\
p\cdot p_u-\bar{P}_{22}-{p}^2& 0 \emat P_{j-\half}'(x_u)\right], \no\\
B_{Pj}&=&\frac{(-1)^{j-\half}(j-\half)!}{(2j)!!}
(\tilde{k}^2 \tilde{k'}^2)^{(j-\half)/2} \,\left[-\bmat 1&0\\
0&0 \emat P_{j+\half}'(x_u) \right. \no\\
&&\left. +\bmat \bar{P}_{22}& m p'\cdot p_u/u\\
m p\cdot p_u/u& 1 \emat P_{j-\half}'(x_u)\right], \no
\eea
where $P_l'(x)$ is the first derivative of the Legendre polynomial, and
\beq
\tk=k-\frac{(k\cdot p_u)}{u} p_u,\,\,\,\,
{\tk'} ={k'}-\frac{(k'\cdot p_u)}{u} p_u,\,\,\,\,
x_u= -\frac{\tk\cdot \tk'}{(\tk^2 \tilde{k'}^2)^{\half}}.
\eeq

\ni
{(c)} The $\De$ exchange using the gauge-invariant coupling \eref{int3},
according to \Eqref{giample} gives
\bea
A^I&=&\frac{f_{\pi N\De}^2}{4\pi m_\pi m}\frac{u}{u-m^2}\, A_{P\thalf}\,
{\cal F}(I),\\
B^I&=&\frac{f_{\pi N\De}^2}{4\pi m_\pi m}\frac{u}{u-m^2}\, B_{P\thalf}\,
{\cal F}(I).
\eea

\subsection{Meson-exchange graphs}
\ni
For the isoscalar meson: ${\cal F}(\half)={\cal F}(\thalf)=2$.

\ni
For the isovector meson: ${\cal F}(\half)=2$, ${\cal F}(\thalf)=-1$.

\ni
{(a)} $\rho$-meson exchange:
\bea
A^I&=&\frac{g_{\rho\pi\pi}g_{\rho NN}}{4\pi}\frac{1}{m_\rho^2-t}\,
\bmat  \frac{\kappa_\rho}{2m_N}( k^2 +{k'}^2- 2s) & -1 \\ 
 -1 & -\frac{\kappa_\rho}{m_N} \emat\, {\cal F}(I),\\
B^I&=&\frac{g_{\rho\pi\pi}g_{\rho NN}}{4\pi}\frac{1}{m_\rho^2-t}\,
\bmat  2 & \frac{\kappa_\rho}{m_N} \\ 
 \frac{\kappa_\rho}{m_N}&0 \emat\, {\cal F}(I).
\eea

\ni
{(b)} $\sigma$-meson exchange:
\bea
A^I&=&\frac{g_{\si\pi\pi}g_{\si NN}}{8\pi m_\pi}\left[
c_\si \frac{m_\pi^2}{m_\si^2} - 
\frac{\half (k^2 +{k'}^2)}{m_\si^2-t}\right]\,
\bmat  1 & 0 \\ 
 0 & 0\emat\, {\cal F}(I),\\
B^I&=& 0.
\eea

\section{\pn amplitudes for higher-spin baryon exchange}

The use of GI couplings of higher-spin baryons \cite{PT99} allows
us to treat exchanges of baryons with any spin in a straightforward
way. Taking the point of view that consistent $\pi NN^\ast_j$ couplings 
(where by $N^\ast_j$ we denote the spin-$j$ baryon, $j\geq 3/2$) 
are those invariant under the appropriate gauge transformations 
of the $N^\ast_j$ field, we can write down the following  {\em ansatz} for
the $\pi N$ scattering amplitude through a spin-$j$ baryon 
tree-level exchange: 
\bea
\eqlab{fjprop}
M(k',k; P) &=& (-1)^l\frac{l!}{(2l+1)!!}\, \frac{g^2}{\slap-m}\no\\ 
&\times& \left[ P_{l+1}' (-\hk\cdot\hk') + \sla{\hk}'\,\sla{\hk}
\, P_{l}' (-\hk\cdot\hk')\right]\,
\left( P^2 \sqrt{\tk^2\, \tilde{k'}^2} \right)^l,
\eea
where $l=j-\half$, $P_l'(x)$ is the first derivative of 
the Legendre polynomial, $P$ is the momentum of the exchanged baryon,
and
\beq
\eqlab{spinmom}
\tk_\mu = k_\mu - \frac{k\cdot P}{P^2}\,P_\mu, \hspace{1cm}
\hk_\mu= \tk_\mu/\sqrt{\tk^2}.
\eeq
This amplitude is actually just the spin-$j$ projection operator
contracted with the external pion momenta and multiplied by 
$g^2 p^{2l} (\slap-m)^{-1}$. Because of the projection operator,
the amplitude contains only the spin-$j$ contributions, as we will
now demonstrate.

In the \CMS\ $P=(W,\vec{0})$, hence $\tk=(0,\vec{k})=(0,-\vec{p})$.
The $s$-channel helicity amplitude is then written as,
\bea
M^{\rho'\rho}_{\la'\la} &=& \bar{u}^{\rho'}_{\la'}(\vp')\,M(k',k; p)\,
u^{\rho}_{\la}(\vp) =
C_j\, \bar{u}^{\rho'}_{\la'}(p')\, 
\frac{W\ga_0 + m}{s - m^2}\,
\left[ P_{l+1}' (x_s) \right.\no\\
&+& \left. (\rp\rp')^{-1}(\vec{\ga}\vec{p}\,')(\vec{\ga}\vec{p})
\, P_{l}' (x_s) \right]\, (\rp \rp' s)^l
u^{\rho}_{\la}(\vp),
\eea
where $x_s =\cos\th,$ $\th$ is the c.m. scattering angle, $s=W^2$,
and the constant factors are absorbed in $C_j$. 
Using the Dirac equation,
$(\vec{\ga}\vec{p})\, u^{\rho}_{\la}(p) = 
 (\rho E_p \ga_0-m_\N)\,u^{\rho}_{\la}(p),$
we obtain
\bea
M^{\rho'\rho}_{\la'\la}&=& C_j\, \bar{u}^{\rho'}_{\la'}(p') \, 
\left[ \frac{W\ga_0 + m}{s - m^2}\,P_{l+1}' (x_s) \right. \nonumber\\
&+& \left. \,\,\frac{-W\ga_0 + m}{s - m^2}\, 
\frac{\rho' E_{p'}\ga_0-m_\N}{\rp'}\frac{\rho E_p\ga_0-m_\N}{\rp}
\, P_{l}' (x_s) \right]\, (\rp \rp' s)^l u^{\rho}_{\la}(p).
\eea
The contribution of this amplitude to the partial wave with total
spin $J$ is given by the corresponding partial-wave amplitude.
The latter can be obtained by the procedure of
Appendix~B, for that we only need to know the following angular
integrals:
\beq
\frac{1}{2}\int_{-1}^1 \!\dd x\,P_{l}'(x)\,P_L(x)
= \left\{ \begin{array}{ll} 1, & L=l-2n-1\\
 0, & {\rm otherwise} \end{array} \right. ,
\eeq
where $n$ should be an integer between 0 and $(l-1)/2$.
The resulting parity-conserving partial-wave amplitude is then
found to be
\bea
M^{\rho'\rho\,J}_r&=& C_j\,(\rp \rp' s)^{l}\times
\left\{ \begin{array}{ll} 
S^{(+r)}\,R^{\rho'\rho}_r,
& J=j,\\
S^{(-r)}\,(R^{\rho'\rho}_{-r} + {\cal N}_{-r}'
{\cal N}_{-r}\,R^{\rho'\rho}_{r}),
& J=j-1-2n,\\
S^{(+r)}\,(R^{\rho'\rho}_r + {\cal N}_{r}'
{\cal N}_{r}\,R^{\rho'\rho}_{-r}),
& J=j-2-2n, \end{array} \right.
\eea
where ${\cal N}_{\pm}= (\rho E_p\pm m_\N)/\rp$,
$S^{(\pm)}=(\pm W-m)^{-1}$, and factors $R$ are defined
in \Eqref{rfactors}. Using the explicit form of $R$,
we can see that the lower partial-wave contributions vanish exactly. 
Thus, amplitude \Eqref{fjprop} has only the highest-spin
contribution:
\beq
\eqlab{pwaj}
M^{\rho'\rho\,J}_r= \de_{jJ}\, 
C_j\,(\rp \rp' s)^{l} S^{(r)}\,R^{\rho'\rho}_r\,.
\eeq

\bibliographystyle{prsty}

\newpage
\epsfxsize=14cm
\begin{figure}[h]
\epsffile{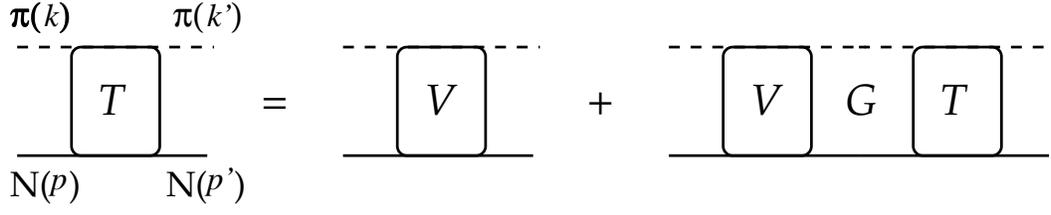}
\caption{ Diagrammatic form of a relativistic $\pi N$ scattering  
equation.}
\figlab{bsef}
\end{figure}

\epsfxsize=14cm
\begin{figure}[t]
\epsffile{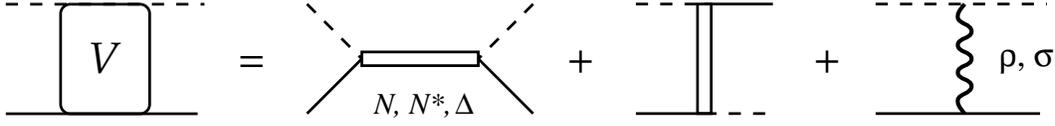}
\caption{ The tree-level $\pi N$  potential.}
\figlab{potf}
\end{figure}

\epsfxsize=14cm
\begin{figure}[p]
\begin{center}
\epsffile{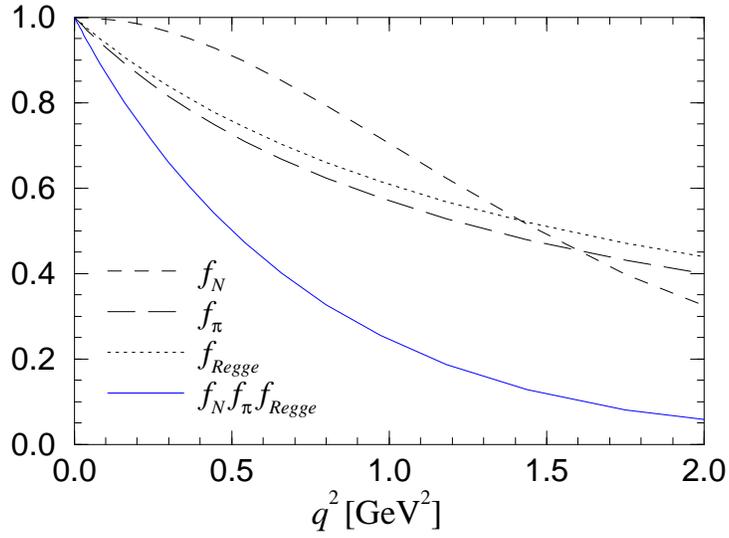}
\caption{ Cutoff form factors as a function of the
off-shell 3-momentum squared.}
\figlab{cform1}
\end{center}
\end{figure}

\epsfxsize=14cm
\begin{figure}[p]
\begin{center}
\epsffile{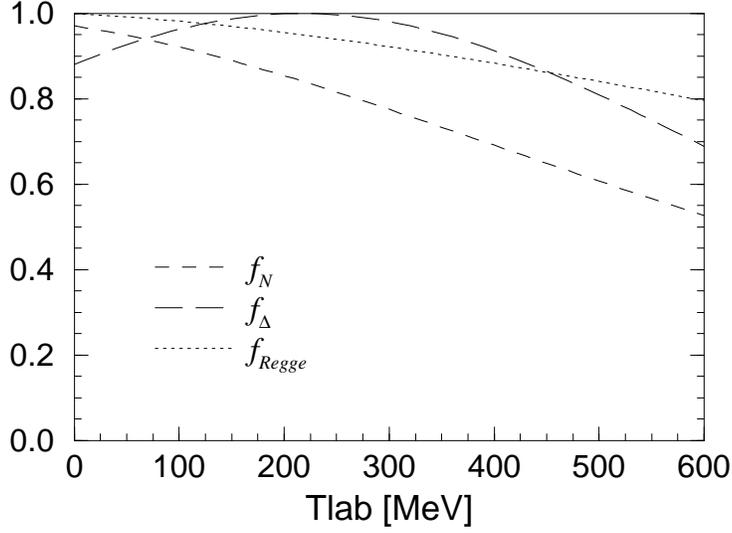}
\caption{ Cutoff form factors as the function of the pion kinetic
energy in the lab.}
\figlab{cform2}
\end{center}
\end{figure}

\begin{figure}[h]
\begin{center}
\epsffile{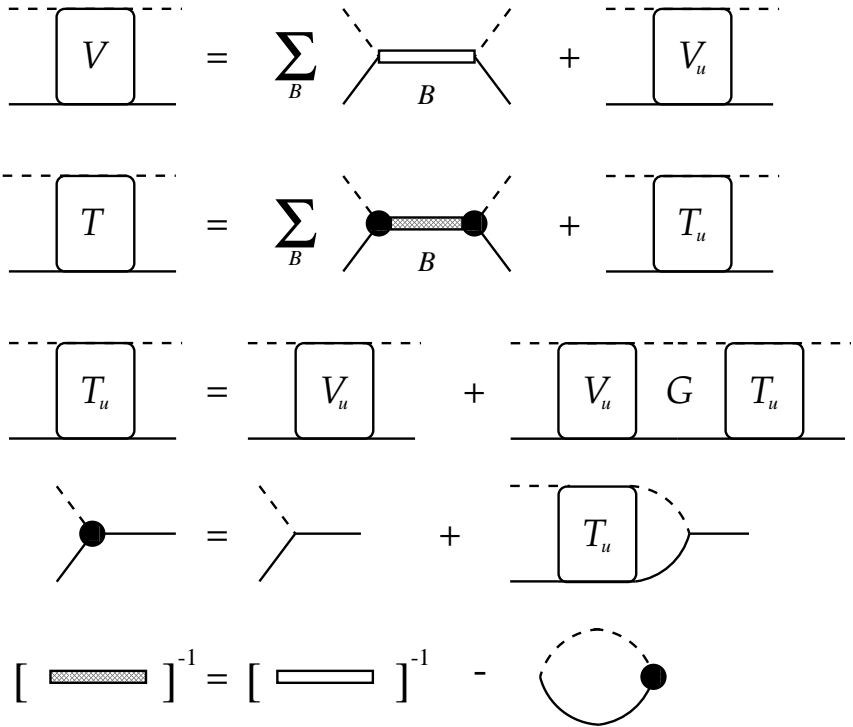}
\end{center}
\medskip
\caption{ Rewriting the equation for the case when the potential 
can be presented as the sum of separable and non-separable potentials.}
\figlab{irred}
\end{figure}

\begin{figure}[p]
\begin{center}
\epsffile{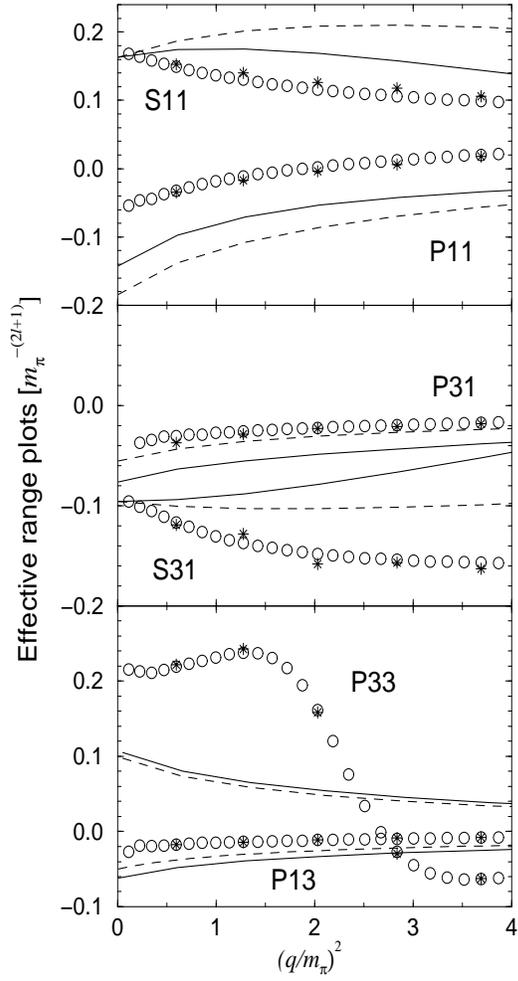}
\caption{Effective range plots for the 
nucleon and $\rho$-meson exchange contribution.
Dashed lines: $\kappa_\rho=0$ (Weinberg-Tomozawa contact term).
 Solid lines: $\kappa_\rho=3.7$ (vector-meson dominance).
The data points are extracted from the KH80 (stars)
and SM95 (dots) partial-wave analyses.}
\figlab{plot0}
\end{center}
\end{figure}

\begin{figure}[p]
\begin{center}
\epsffile{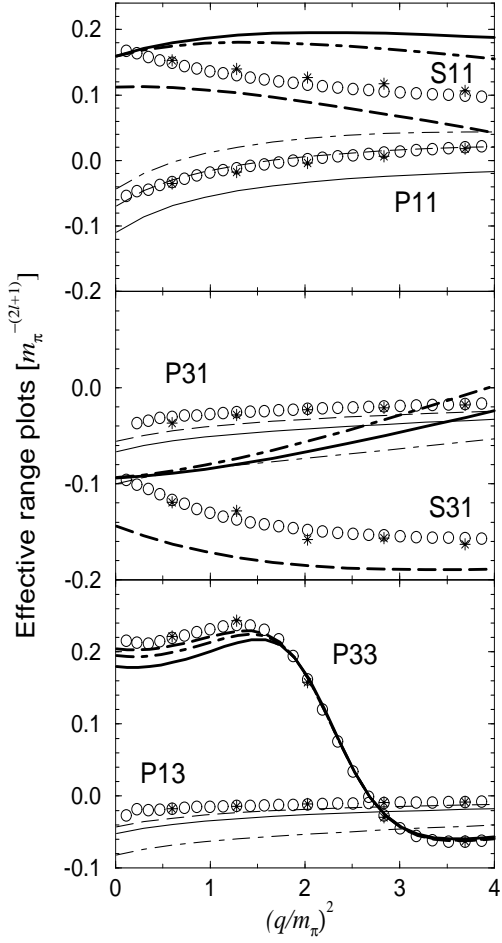}
\caption{Effective range plots for the
sum of the $N$, $\De$ and $\rho$ exchanges with
$g_{\pi NN}^2/4\pi=13.6$, $g_{\rho}^2/4\pi=3$,
$\kappa_\rho = 3.7$, $f_{\pi N\De}^2/4\pi=0.36$. 
Dashed lines: Peccei choice; dashed-dotted lines:
NEK choice; solid lines: gauge-invariant $\pi N\De$ coupling. 
Results for $S_{11}$, $S_{31}$, and $P_{33}$ waves are shown
in bold lines. Data points are same as in \Figref{plot0}.}
\figlab{plot1}
\end{center}
\end{figure}

\epsfxsize=15cm
\begin{figure}[h]
\begin{center}
\epsffile{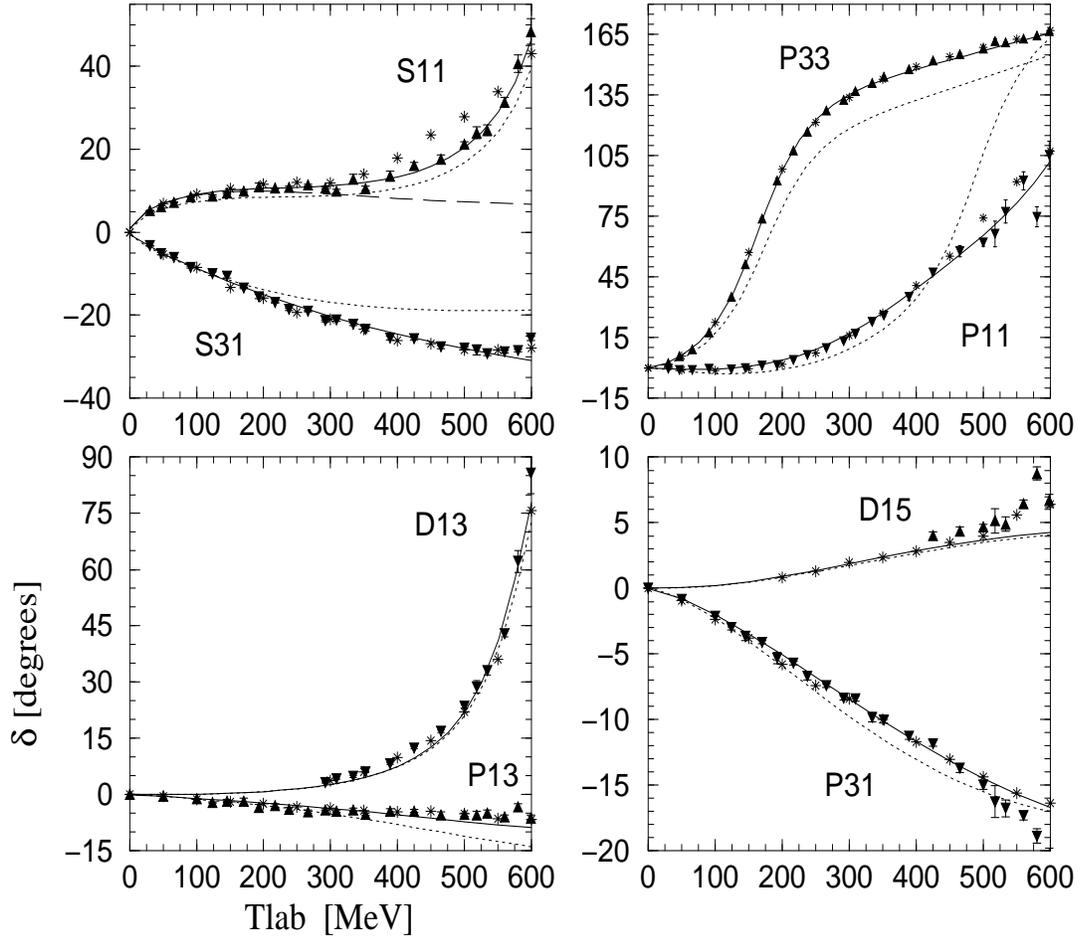}
\caption{ The description of the $S$- and $P$- and $D$-wave $\pi N$ 
phase-shifts up to
600 MeV.  Data points are from the SM95 (triangles) and KH80 (stars)
partial-wave analysis. Solid lines represent the model solution. 
Dotted lines represent the calculation where the principal value
part of the rescattering integrals is switched off
(\ie the K-matrix approximation with the same set of
parameters). Dashed line for the $S_{11}$ shows the calculation
when the pole contribution of the $S_{11}$ resonance is switched off.}
\figlab{pin600f}
\end{center}
\end{figure}

\epsfxsize=15.0cm
\begin{figure}[h]
\begin{center}
\epsffile{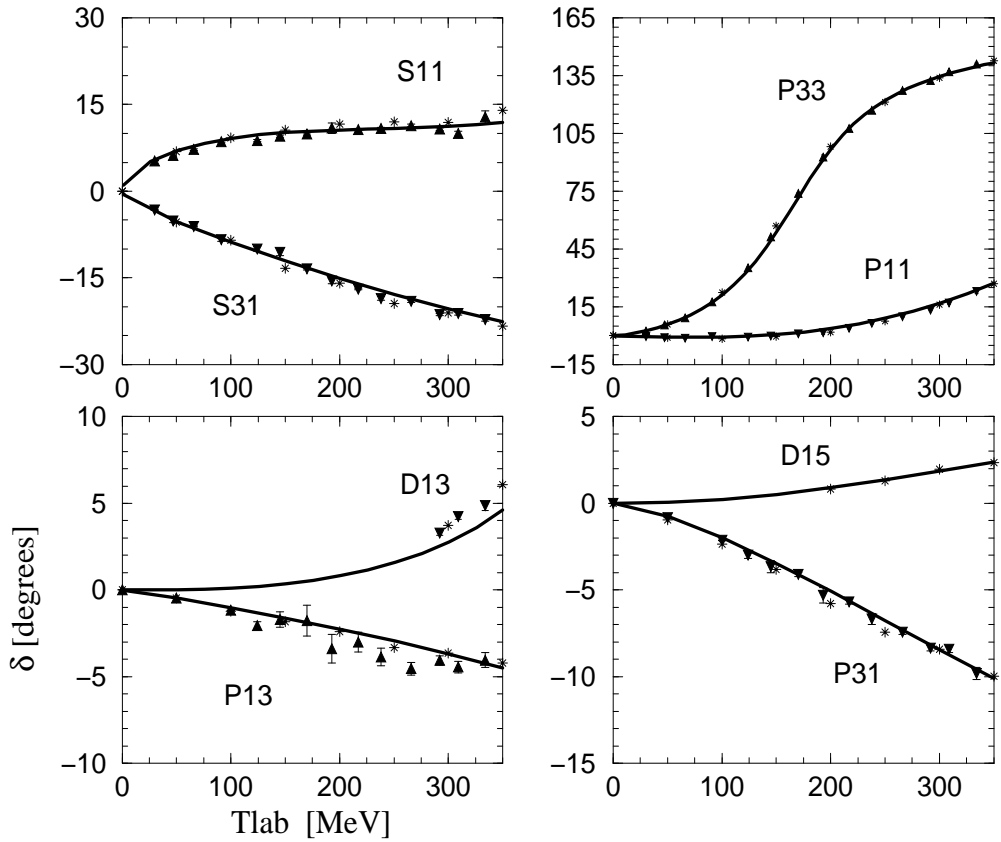}
\end{center}
\caption{ The description of $\pi N$ phase-shifts
up to 350 MeV pion lab kinetic energy.  
Solid lines represent the full model calculation.
Legend for the data points 
is the same as in \Figref{pin600f}.} 
\figlab{pin350f}
\end{figure}

\begin{figure}[h]
\begin{center}
\epsffile{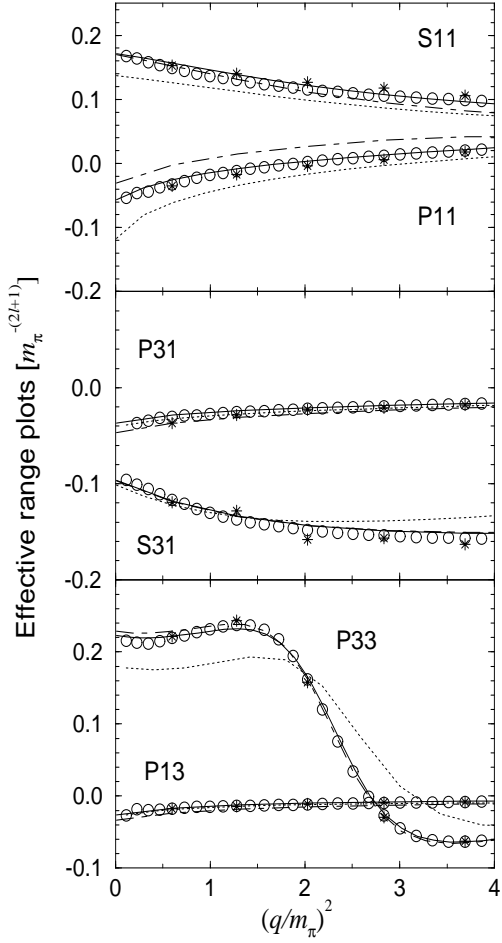}
\caption{ Description of the effective range plots for the
$S$- and $P$-waves.  Solid lines represent model calculation.
Dotted lines are the $K$-matrix approximation with the same set
parameters. Dashed-dotted line represent the full calculation
with $\kappa_\rho=3.7$.}
\figlab{effr}
\end{center}
\end{figure}

\epsfxsize=15.0cm
\begin{figure}[h]
\begin{center}
\epsffile{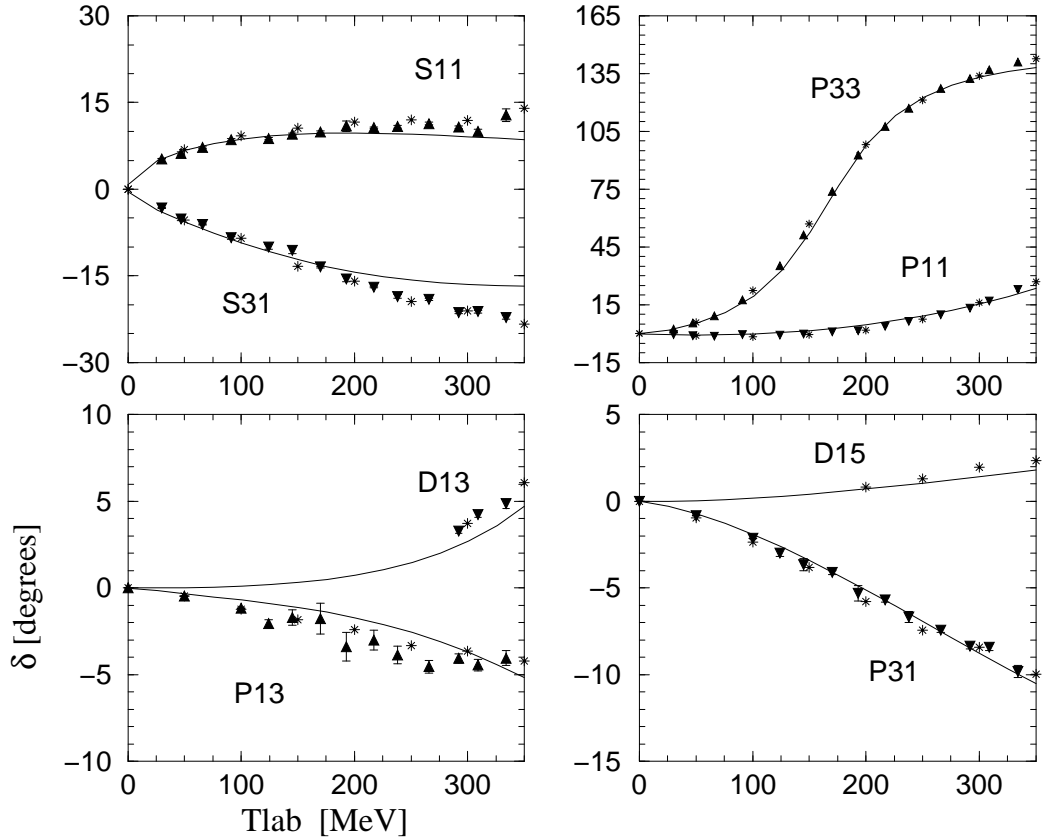}
\end{center}
\caption{ Fit of the $K$-matrix approximated model to
$\pi N$ phase-shifts
up to 350 MeV pion lab kinetic energy.} 
\figlab{kmat350f}
\end{figure}

\epsfxsize=11cm
\begin{figure}[h]
\begin{center}
\epsffile{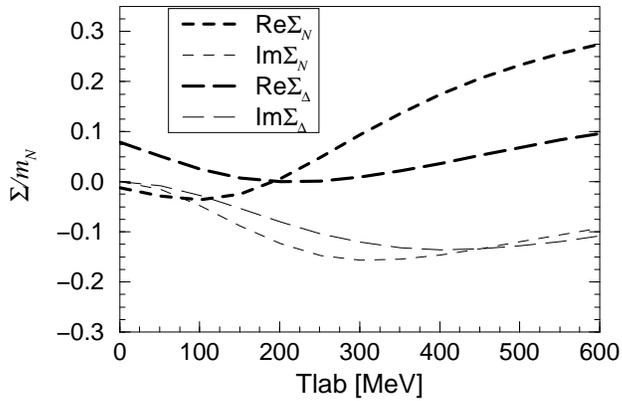}
\caption{ The real and imaginary parts of the self-energies}
\figlab{sef}
\end{center}
\end{figure}

\epsfxsize=11cm
\begin{figure}[h]
\begin{center}
\epsffile{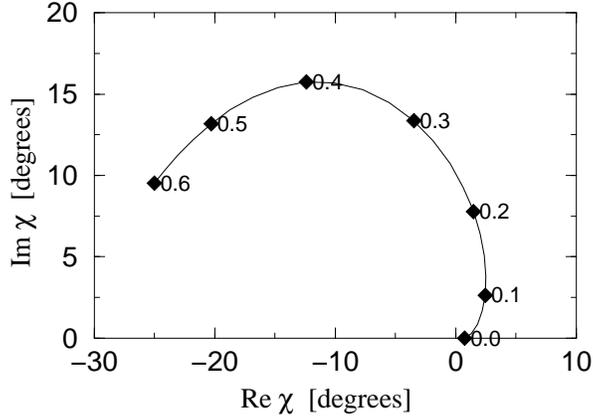}
\end{center}
\caption{ The energy dependence of the 
$N N^\ast$ mixing angle. The numbers near the curve stand for the
corresponding value of the pion lab kinetic energy in GeV.}
\figlab{nnmixf}
\end{figure}

\epsfxsize=13cm
\begin{figure}[h]
\begin{center}
\epsffile{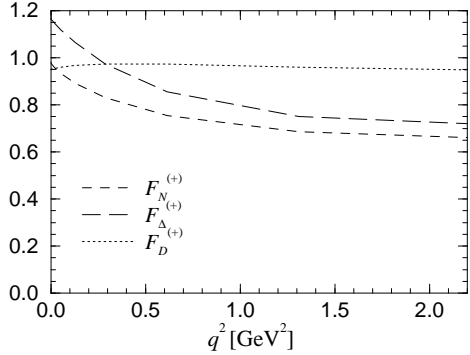}
\caption{ The dynamical $\pi NN$, $\pi N\De$ and
 $\pi ND_{13}$ form factors, as a function of the
off-shell 3-momentum squared of the $\pi N$ state. The energy of the
baryon is fixed at the $\pi N$ threshold.}
\figlab{form1}
\end{center}
\end{figure}
\epsfxsize=13cm
\begin{figure}[h]
\begin{center}
\epsffile{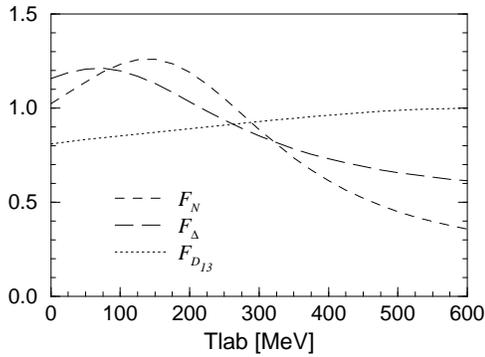}
\caption{ The dynamical half-off-shell form factors 
for the case when the pion and nucleon are on-shell, as a function
of the pion lab kinetic energy.}
\figlab{form2}
\end{center}
\end{figure}


\ecol
\begin{table}[t]
\begin{center}
\begin{tabular}{crrrrrr|rrr} 
$l_{2I\, 2J}$ & $N\rho$ (WT) & $N\rho$ (VMD)& Peccei 
& NEK & $N\rho\De$ (WT) & $ N\rho\De $ (VMD)
& KH80 \cite{KH80} & KA86 \cite{KA86} & SM95 \cite{ASW95}\\ \hline
S$_{11}$ &  0.171    &  0.171   & 0.144    & 0.171     & 0.171 & 0.171 
& 0.173 & 0.175& 0.172\\
S$_{31}$ &  $-0.100$ & $-0.100$ & $-0.150$ & $-0.100$  & $-0.100$ & $-0.100$
& --0.101 & --0.100 & --0.097\\
P$_{11}$ &  $-0.185$ & $-0.145$ & $-0.077$ & $-0.076$  & $-0.127$ & $-0.092$
& --0.081 & --0.078 & --0.068\\
P$_{31}$ &  $-0.059$ & $-0.079$ & $-0.064$ & $-0.085$  & $-0.043$ & $-0.066$
& --0.045 & --0.043 & --0.040\\
P$_{13}$ &  $-0.047$ & $-0.064$ & $-0.049$ & $-0.069$  & $-0.032$ & $-0.050$
& --0.030 & --0.030 & --0.021\\
P$_{33}$ &  0.101    & 0.109    & 0.194 & 0.197     & 0.224    &0.222
& 0.214 & 0.215 & 0.209\\ \hline
\\
$\chi^2$ &  177      & 129      & 35      & 31        & 26       & 15
& 2.1 & 1.6 & $\times$\\
\end{tabular}
\end{center}
\caption{  The $S$- and $P$-wave $\pi N$ scattering lengths in
different models are compared to the partial-wave analyses.
The $\chi^2 $ represent the chisquare value per point with respect to the
SM95 analysis (the analysis error is taken to be the same
for all points and is equal to 0.005).   Parameters
corresponding to different calculations are given in Table~\tabref{cc1}.}
\tablab{sl1}
\end{table}

\begin{table}[t]
\begin{center}
\begin{tabular}{ccccccc} 
parameters & $N\rho$ (WT) & $N\rho$ (VMD)& Peccei 
& NEK & $N\rho\De$ (WT) & $ N\rho\De $ (VMD)
\\ \hline
\\
$g_{\pi NN}^2/4\pi$ &  13.8  &  13.8  & 13.8    & 13.8    & 13.8 & 13.8 
\\
$g_{\rho NN}g_{\rho\pi\pi}/4\pi $ & \bf 3.15 & \bf 3.15
& \bf 3.4 & \bf 3.15  & \bf 3.15 & \bf 3.15
\\
$\kappa_\rho$ &  0.0 & 3.7 & 3.7 & 0.0  
& 0.0 & 3.7
\\
$f_{\pi N\De}^2/4\pi $ & $\times$  & $\times$ & \bf 0.31 & \bf 0.36  
& \bf 0.6 & \bf 0.55
\\
$z_\De$ & $\times$  & $\times$ & $-0.25$ & 0.5  & $\times$ & $\times$
\\
\end{tabular}
\end{center}
\caption{Parameters corresponding to calculations of Table~\tabref{sl1}.
Values in bold were varied to give a best fit of the scattering lengths.}
\tablab{cc1}
\end{table}

\becol
\begin{table}[t]
\begin{center}
\begin{tabular}{crr|r} 
$l_{2I\, 2J}$  & $N\rho\De\si$ (A) & $ N\rho\De\si $ (B) 
&  SM95 \cite{ASW95}\\ \hline
S$_{11}$ &  0.170   &  0.170   &   0.172\\
S$_{31}$ &  $-0.100$ & $-0.100$ &  $-0.097$\\
P$_{11}$ &  $-0.068$ & $-0.069$ &   $-0.068$\\
P$_{31}$ &  $-0.039$ & $-0.039$ &  $-0.040$\\
P$_{13}$ &  $-0.024$ & $-0.023$ &   $-0.021$\\
P$_{33}$ &  0.209    & 0.209    &   0.209\\ \hline
\\
$\chi^2$ &  0.16      & 0.13     &  $\times$\\
\end{tabular}
\end{center}
\caption{ Scattering lengths resulting from the sum of
$N$, $\De$, $\rho$, and $\si$ tree-level exchanges. Parameters
corresponding to different models are given in Table~\tabref{cc2}.}
\tablab{sl2}
\end{table}

\begin{table}[t]
\begin{center}
\begin{tabular}{cccc} 
parameters  & $N\rho\De\si$ (A) & $ N\rho\De\si $ (B) 
\\ \hline
\\
$g_{\pi NN}^2/4\pi$ &  13.8  &  13.8  
\\
$g_{\rho NN}g_{\rho\pi\pi}/4\pi $ & \bf 3.15 & \bf 3.15 
\\
$\kappa_\rho$ &  \bf 0.0 & \bf 4.6
\\
$f_{\pi N\De}^2/4\pi $ & \bf 0.28   & \bf 0.30 
\\
$z_\De$  & \bf 0.5 & $\times$  
\\
$g_{\si NN}g_{\si\pi\pi} m_N/(8\pi m_{\pi})$ & \bf 2.1 & \bf 2.0
\\
$c_\si$ &  1.0 & 1.0 
\end{tabular}
\end{center}
\caption{Parameters corresponding to the calculation in 
Table~\tabref{sl2}.
Values in bold were established by the fit. Model A uses the
conventional $\pi N\De$ coupling, while model B uses the
``gauge-invariant'' coupling.}
\tablab{cc2}
\end{table}

\begin{table}[p]
\begin{center}
\begin{tabular}{c|l|l}
field& {\rm coupling constants} & {\rm masses [GeV]}\\ \hline
$N$ &$\frac{g_{\pi NN}^2}{4\pi} = 13.8$ ($f_{\pi NN}^2$=0.0757) 
& $m_N = 0.9383,\,\, \La_N={\bf 1.8}$\\
$\pi$& & $m_\pi=0.139,\,\, \La_\pi={\bf 1.16}$\\
$N^\ast$& $\frac{g_{\pi NN^\ast}^2}{4\pi} ={\bf 2.7}$  
& $m_{N^\ast} ={\bf 1.438},\,\, \La_{N^\ast}=\La_N$\\
$\De$& $\frac{f_{\pi N\De}^2}{4\pi} ={\bf 0.43}$ 
& $m_\De={\bf 1.252},\,\,\La_\De={\bf 1.5}$\\
$D_{13}$& $\frac{f_{\pi N D}^2}{4\pi} ={\bf 0.6}$ & 
$m_D={\bf 1.525},\,\,\La_D={\bf 1.7}$\\
$\rho$& $\frac{g_{\rho NN}g_{\rho\pi\pi}}{4\pi} ={\bf 2.85},\, 
\kappa_\rho={\bf 1.8}
$& $m_\rho={0.77},\,\,\La_\rho={\bf 1.9} $\\
$\si$& $\frac{g_{\si NN}g_{\si\pi\pi}}{4\pi} \frac{m_N}{2m_\pi} 
= {\bf 1.0},\,
c_\si={\bf 0.75}$ & $m_\si=0.55,\,\, \La_\si={\bf 1.1}$ \\
$S_{11}$& $\frac{f_{\pi N S_{11}}^2}{4\pi} ={\bf 0.6}$ & 
$m_{S_{11}}={\bf 1.555},
\,\,\La_{S_{11}}={\bf 2.0}$\\
\end{tabular}
\end{center}
\caption{ The model parameters. The values given in bold were varied 
during the fit.}
\tablab{modelpar}
\end{table}

\begin{table}[t]
\begin{center}
\begin{tabular}{c|l|l}
field & {\rm coupling constants} & {\rm masses [GeV]}\\ \hline
$N$ &$\frac{g_{\pi NN}^2}{4\pi} = {13.8}$ ($f_{\pi NN}^2$=0.0757) 
& $m_N = {0.9383},\,\, \La_N={\bf 2.2}$\\
$\pi$& & $m_\pi={0.139}$\\
$N^\ast$& $\frac{g_{\pi NN^\ast}^2}{4\pi} = {\bf 2.6}$  
& $m_{N^\ast} = {\bf 1.47},\,\, \La_{N^\ast}=\La_N$\\
$\De$& $\frac{f_{\pi N\De}^2}{4\pi} = {\bf 0.36}$ & 
$m_\De={\bf 1.232},\,\,\La_\De={\bf 2.6}$\\
$D_{13}$& $\frac{f_{\pi N D}^2}{4\pi} = {\bf 0.8}$ &
 $m_D={\bf 1.525},\,\,\La_D={\bf 1.7}$\\
$\rho$& $\frac{g_{\rho NN}g_{\rho\pi\pi}}{4\pi} = {\bf 3.1},\, 
\kappa_\rho={\bf 2.7}
$& $m_\rho={0.77},\,\,\La_\rho={\bf 1.3} $\\
$\si$& $\frac{g_{\si NN}g_{\si\pi\pi}}{4\pi} \frac{m_N}{2m_\pi} = {\bf 1.3},\,
c_\si={\bf 1.0}$ & $m_\si=0.55,\,\, \La_\si={\bf 1.9}$ \\
\end{tabular}
\end{center}
\caption{ The parameters of the $K$-matrix calculation, see 
\Figref{kmat350f}.}
\tablab{kmatpar}
\end{table}

\begin{table}[b]
\begin{center}
\begin{tabular}{c|lll}
field& $Z_1$ & $Z_2$ & $m_0$ [GeV] \\ \hline
$N$ & 0.63 & 0.72 & 1.090\\
$N^\ast$&0.63 & 0.95 & 0.960\\
$\De$& 0.68& 1.01 & 1.390 \\
$D_{13}$& 0.93 & 1.03 & 1.512\\
$S_{11}$& 0.97& 1.09 & 1.555\\
\end{tabular}
\end{center}
\caption{ The renormalization constants obtained in the model.}
\tablab{renpar}
\end{table}

\ecol

\end{document}